%% file: AGBbin11.tex
\documentclass[fleqn,usenatbib]{mnras} 

\usepackage{newtxtext,newtxmath}

\usepackage[T1]{fontenc}
\usepackage{lipsum}
\usepackage[normalem]{ulem}
\usepackage{xcolor}
\usepackage{physics}
\usepackage{bm}
\usepackage{graphicx}	
\usepackage{amsmath}	
\usepackage{multicol}
\usepackage{caption}
\usepackage{subcaption}
\usepackage{orcidlink}
\usepackage{float}

\DeclareRobustCommand{\VAN}[3]{#2}
\let\VANthebibliography\thebibliography
\def\thebibliography{\DeclareRobustCommand{\VAN}[3]{##3}\VANthebibliography}

\newcommand{\Msun}{M$_{\odot}$} 
\newcommand{\Rsun}{R$_{\odot}$} 
\newcommand{\Lsun}{L$_{\odot}$} 
\newcommand{\kapunits}{~cm$^2$~g$^{-1}$} 
\newcommand{\phant}{{\sc phantom}}

\newcommand{\mesa}{{\sc mesa}}

\newcommand{\review}{}
\newcommand{\secondreview}{}

\title[Common envelope simulations with dust I.]{Dust Formation in Common Envelope Binary Interaction --- I:\\ 3D Simulations Using the Bowen Approximation}

\author[M. González-Bolívar et al.]{Miguel González-Bolívar \orcidlink{0000-0002-5939-9269 }$^{1,2}$ \thanks{E-mail: miguel-angel.gonzalez-boliv@hdr.mq.edu.au},
Orsola De Marco\orcidlink{0000-0002-1126-869X}$^{1,2}$,
\newauthor
Luis C. Bermúdez-Bustamante\orcidlink{0000-0002-3629-6259 }$^{1,2}$ 
Lionel Siess\orcidlink{0000-0001-6008-1103}$^3$ and Daniel J. Price\orcidlink{0000-0002-4716-4235}$^4$
\\
$^{1}$ School of Mathematical and Physical Sciences, Macquarie University, Sydney, NSW 2109, Australia\\
$^{2}$ Astronomy, Astrophysics and Astrophotonics Research Centre, Macquarie University, Sydney, NSW 2109, Australia\\
$^{3}$Institut d’Astronomie et d’Astrophysique, Université Libre de Bruxelles (ULB), CP 226, 1050 Brussels, Belgium\\
$^{4}$School of Physics and Astronomy, Monash University, Vic. 3800, Australia\\
}

\date{Accepted XXX. Received YYY; in original form ZZZ}

\pubyear{2023}

\begin{document}
\label{firstpage}
\pagerange{\pageref{firstpage}--\pageref{lastpage}}
\maketitle

\begin{abstract}
We carried out 3D smoothed particle hydrodynamics simulations of common envelope binary interaction using the approximation of Bowen to calculate the dust opacity in order to investigate the resulting dust-driven accelerations. We have simulated two types of binary star: a 1.7 and a 3.7 \Msun thermally-pulsating, asymptotic giant branch stars with 0.6 \Msun companions. We carried out simulations using both an ideal gas and a tabulated equation of state, with the latter considering the recombination energy of the envelope. We found that the dust-driven wind leads to a relatively small increase in the unbound gas, with the effect being smaller for the tabulated equation of state simulations. Dust acceleration does contribute to envelope expansion with only a slightly elongated morphology, if we believe the results from the tabulated equation of state as more reliable. The Bowen opacities in the outer envelopes of the two models, at late times, are large enough that the photosphere of the post-inspiral object is about ten times larger compared to the same without accounting for the dust opacities. As such, the prediction of the appearance of the transient would change substantially if dust is included.
\end{abstract}


\input{introduction/introduction}

\input{setup/setup}
\input{results/results}

\input{discussion/discussion_conclusions}

\section*{Acknowledgements}

MGB acknowledges funding support from Macquarie University through the International Macquarie University Research Excellence Scholarship (iMQRES), to Ryosuke Hirai, Mike Lau and Ilya Mandel for his insightful commentaries during the course of this research, as well as to the Universidad Autónoma de Ciudad Juárez. OD and LCBB acknowledge funding from Australian Research Council Discovery Project, DP210101094. DP acknowledges ARC funding via DP180104235.  LS is a senior research associate from F.R.S.- FNRS (Belgium). Parts of this research work were performed on the Gadi supercomputer of the National Computational Infrastructure (NCI), supported by the Australian Government, and by Oracle Cloud credits and related resources provided by Oracle for Research. This research has made use of the NASA’s Astrophysics Data System (ADS).

\section{Data availability}

\review{Initial \phant\ snapshot and .in files and can be accessed in \url{https://zenodo.org/records/10052253}.} The data underlying this article will be shared on reasonable request to the corresponding author. 



\bibliographystyle{mnras}
\bibliography{biblio,bibliography} 

\appendix

\input{appendices/stellar_relaxation}
\input{appendices/computational_resources}

\bsp	
\label{lastpage}
\end{document}

%% file: introduction/introduction.tex
\section{Introduction}

Common envelope (CE) interaction occurs in binary stars when one of the components' envelope expands beyond its Roche lobe and engulfs its companion. This phase lasts of the order of a dynamical time and ends with either the ejection of the envelope and the shrinkage of the orbit or with partial envelope unbinding and a merger of the stellar core and the companion \citep{Ivanova2013}. CE evolution leads to the formation of compact short-period binaries such as symbiotic binaries, X-ray binaries, cataclysmic variables, double white dwarfs, and at least 20~per cent of all planetary nebulae \citep{DeMarco2017,Jacoby2021}. CE interaction mergers and ejections are also likely to be responsible for a number of intermediate luminosity transients \citep[e.g.,][]{Blagorodnova2017} such as luminous red novae (LRNe). 

Observations show that close to the outburst itself, several LRNe may produce dust, which obscures some of the phases of the event. For example, the LRN V~1309~Sco was a 1--3~{\Msun} contact-binary merger that happened to have been monitored both before and during the outburst and it showed that  dust formed right before the main outburst \citep{Tylenda2011,Nicholls2013}. It is likely that dust is an active ingredient of these binary interactions and outbursts. If dust forms in CE interactions there is a chance that it could participate in the dynamics of the envelope.

Using recombination energy in CE simulations has allowed the question of how the envelope is ejected to be at least partly settled. After an initial injection of orbital energy, it is the liberation of recombination energy that provides an additional source of work \citep[e.g.,][]{Ivanova2015,Ohlmann2016,Reichardt2020,GonzalezBolivar2022,Lau2022a}. The question of whether the recombination photons are captured and do work or stream out of the star is also partly settled \citep{Ivanova2018,Soker2018} with simulations showing substantial energy is deposited where it is likely to be thermalised locally \citep{Reichardt2020,Lau2022a}. What is not completely understood is exactly in what parameter space recombination energy fully unbinds the envelope and in which cases it would be more ineffective. It is therefore possible that dust driving may play a role in the driving of the envelope.

For the case of CE interactions with low- and intermediate-mass stars such as red giant branch (RGB) and AGB stars this is a particularly fitting question because we know that dust formation and dust driven winds, in the case of AGB stars, are the primary mechanism for envelope ejection in single stars. A CE interaction provides a great opportunity for dust to form, because even in the fast dynamical phase, expansion and cooling take place. 
Even if it did not provide a critical ejection-driving mechanism, even small amounts of dust will impact the optical properties of the material and hence the prediction of what such an event will look like.
  
\citet{Lu2013} performed 1D simulations to determine the amount of dust mass produced in CE interactions that happen at the base or tip of the RGB for a range of stellar masses from 1 to 7~\Msun. They considered only oxygen-rich dust types and assumed relationships between the expanding CE radius and the corresponding temperature and density. They concluded that for the top-of-the-RGB systems with the steepest dependence of temperature on radius the dust production from CE interactions could rival that from single stars. 

\cite{Glanz2018} showed analytically based on post-processing by \cite{Passy2012} in their 3D hydrodynamic simulation, that envelopes from red giant stars develop AGB-like conditions in the ejected material, which are suitable for dust condensation, and that CE ejection introduces the conditions for a long-term slow mass-loss phase via gas-dust coupling. This would eject the CE over longer timescales similar to those of single AGB stars ($10^5$~yr). As suggested by \citet{Reichardt2020}, they stated that the high opacities developed early on in the outer CE layers could assist in trapping recombination energy into the envelope. 

\cite{Iaconi2019,Iaconi2020} carried out carbon- and oxygen-dust nucleation simulations based on post-processing the 3D hydrodynamic simulations of a CE between a 0.88~\Msun, 90~\Rsun\ RGB star and a 0.6~\Msun\ companion of \citet{Passy2012}. They concluded that dust formation occurs in the outer layers of the envelope, where most of the envelope mass resides at the end of the simulation. The dust masses they obtained for this simulation are 2 and 2.5$~\times 10^{-3}$~\Msun~yr$^{-1}$ for oxygen and carbon-based dust, respectively, about a third of the value obtained by \citet{Lu2013} for the most productive, top of the RGB models. 

\cite{Bermudez2020} implemented dust-driven winds modelled following the approximation of \citet{Bowen1988} for a 2.2~\Msun, 316~\Rsun\ AGB star orbited by a 0.6~\Msun\ companion at Roche lobe overflow distance, including stellar pulsations and radiation pressure on dust grains. They found effective dust driving where the ejection was predominantly along the equatorial direction. Finally \cite{Siess2022} implemented dust opacities using both the \citet{Bowen1988} implementation and full nucleation in the 3D hydrodynamics code \phant\ \citep{Price2018} and carried out test solutions to verify and validate the approach on single AGB stars.  

In this first of two papers we explore the dust formation properties and dust driving in CE simulations using the approach of \citet{Siess2022}. We carry out two common envelope simulations with a 1.7~\Msun{} and a 3.7~\Msun{} star. The former is the same star used by \citet{GonzalezBolivar2022}, while the latter is a new simulation. In this paper (Paper~I) we only use the simple \cite{Bowen1988} prescription for the dust opacity entering the radiative acceleration. The radiative driving force is proportional to the dust opacity which is itself calculated with an analytical prescription. This approximation is extremely simple and does not increase the computational time of the simulation. 

In Paper~II (Bermudez-Bustamante et al.) the dust opacity will be calculated using the theory of moments \citep{Gail2013}, which is far more costly (the calculation of the acceleration is carried out exactly as in this paper). The aim of this paper is therefore to assess the impact of the opacity calculation in the Bowen approximation. 

In Section~\ref{sec:setup} we describe the simulation setup and give details of the Bowen approximation. Results are described in \ref{sec:results}; we present our  analysis of the orbital evolution and mass unbound in Section~\ref{ssec:orbital_evolution}; we describe the mass and velocity of the gas driven by the dust acceleration in Section~\ref{sec:mvgas} and we analyse the opacity distribution in Section~\ref{ssec:opacity_distribution}. Finally, we show our conclusion in Section~\ref{sec:conclusion}. 


%% file: setup/setup.tex
\section{Setup}
\label{sec:setup}
\subsection{Single and binary star setup}
\label{sec:bin_setup}

We have calculated two stellar models using the 1D implicit code \mesa,\ version 12778, \citep{Paxton2011,Paxton2013,Paxton2015} \review{and evolved them until they reach the thermally-pulsating asymptotic giant branch phase (TP-AGB)}. The first model, \review{used previously by \cite{GonzalezBolivar2022},} is that of a $2$~\Msun\ at the zero age main sequence (ZAMS) with a solar metallicity \citep{Grevesse1998}. This model was evolved until the seventh thermal pulse in the late AGB and reached a mass of 1.7~\Msun, a core mass of $M_c=0.56$~\Msun, a radius of 260~\Rsun\ and an effective temperature of 3227~K. \review{The original \mesa\ stellar profile at the moment of mapping into \phant\ has stellar core radius of 0.01\Rsun}. The stellar wind parameter had the code default values: cool wind RGB scheme was set to ‘Reimers’, with a Reimers scaling factor of 0.1; the cool wind AGB scheme was ‘Blocker’ with a scaling factor of 0.5 and the RGB to AGB wind switch was set at a core helium mass fraction $Y = 10^{-4}$. 

\begin{figure*}
   \centering
         \includegraphics[width=0.43\textwidth]{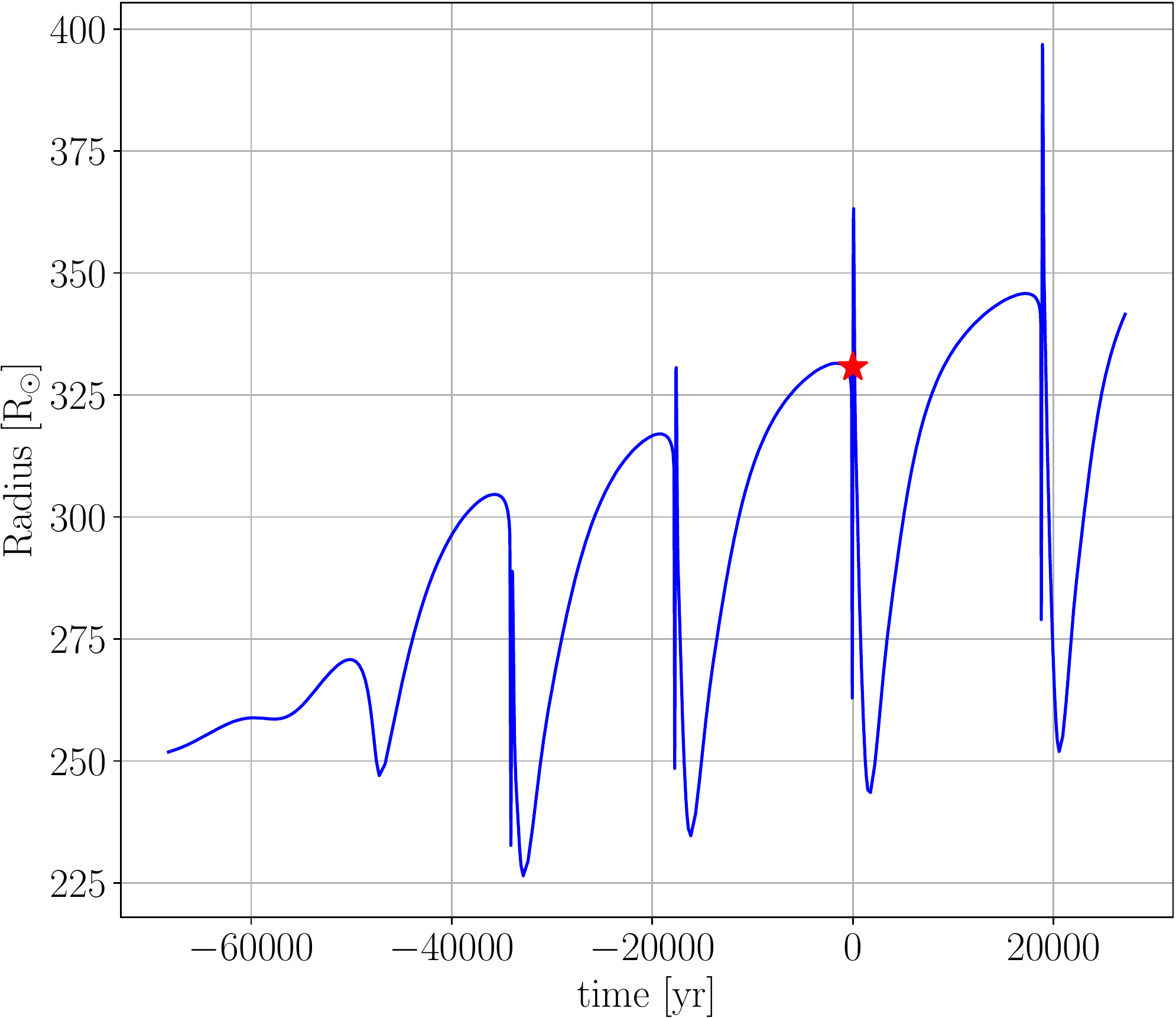}\ \ \ \ \ \ \ \ \ \ \ \ 
         \includegraphics[width=0.43\textwidth]{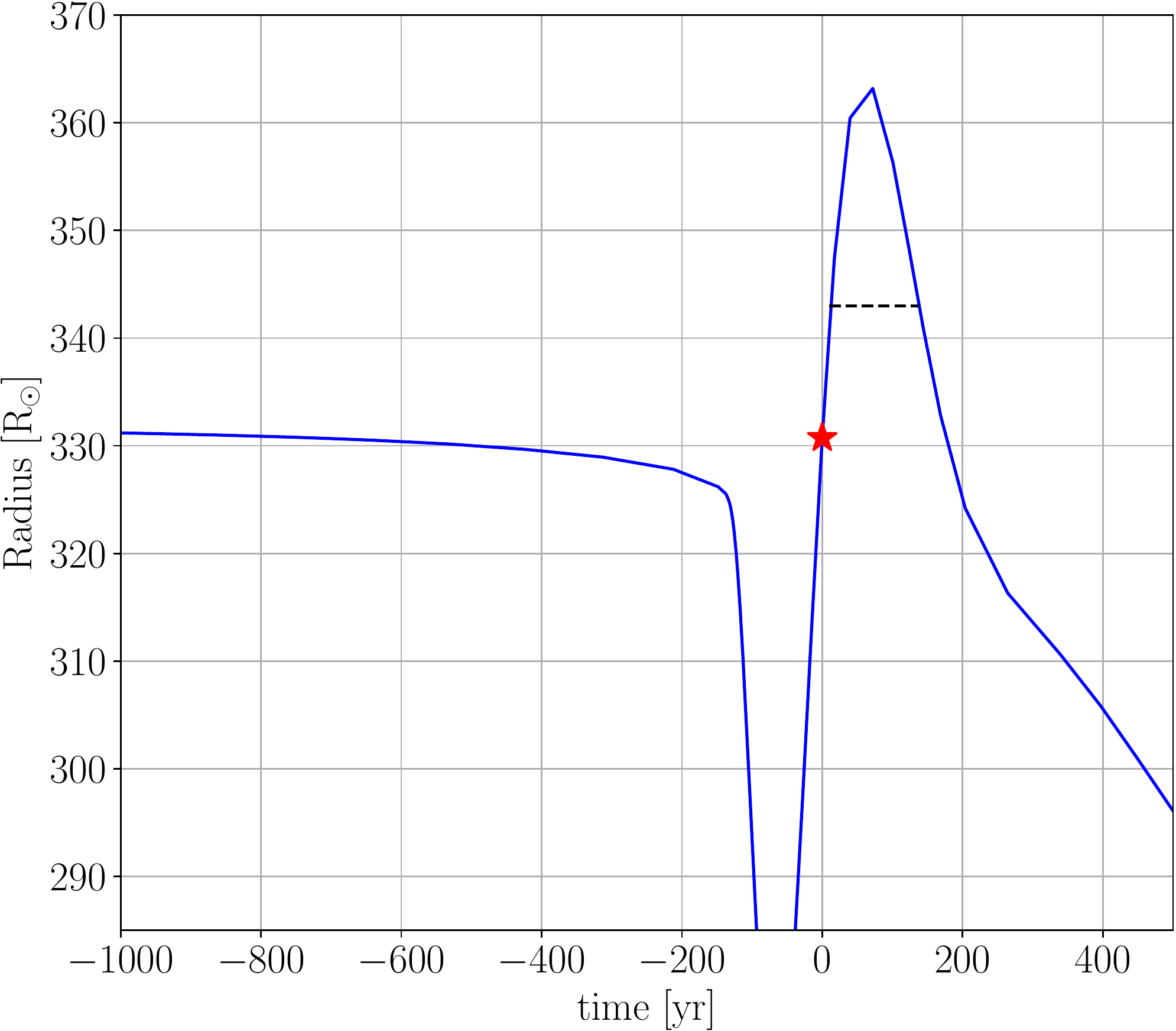}
         \caption{Radial evolution of the 4~\Msun\ (at ZAMS) model during the TP-AGB phase. Curves are shifted so that $t=0$ corresponds to the moment of the evolution in which the model is mapped into the 3D domain in \phant, and is indicated with a red star symbol. Left panel displays the first four thermal pulses. Right panel shows in detail the third pulse. Black dashed line indicates the Roche radius (343~\Rsun) of the donor star in the binary system set for the CE simulations with this model and has a represents an interval of 128 years.}
    \label{fig:4M_thermal-pulse}
\end{figure*}

The second model comes from a $4$~\Msun\ star at the ZAMS that was evolved until the \review{third} thermal pulse and had a mass of 3.7~\Msun, a core mass of $M_c=0.72$~\Msun, a stellar radius of \review{330~\Rsun}, an effective temperature of 3317~K, with the same solar metallicity and mass loss recipes  as the 2~\Msun\ stellar model. The C/O number ratios for both models at the time when they were stopped was 0.32. \review{The wind parameters have the same values as the previous model. The specific selection of this 1D stellar profile was to ensure that its radius value was larger than at any previous moment in the evolution of the star. In this way we can assume that the Roche lobe with the companion could not have been filled before (Figure~\ref{fig:4M_thermal-pulse}). This is the same argument we have used in selecting the 1.7~\Msun\ model in  \citet{GonzalezBolivar2022}.}

These stellar profiles were then mapped in the 3D computational domain using the smoothed particle hydrodynamics (SPH; \citealt{Lucy1977,GingoldMonaghan1977,Monaghan1992,Price2012}) code \phant\ \citep{Price2018} to simulate the CE interaction. We used 1.37 million SPH particles. This is as high a resolution as we have ever attained with \phant\ in CE simulations \citep{Reichardt2019,GonzalezBolivar2022}. Contrary to previous work, here we do not carry out calculations with lower resolution because it is difficult to obtain stable stellar structures with a tabulated equation of state (EoS) with resolutions much lower than 1 million SPH particles. 

\review{Following the mapping of the 1D stellar structure quantities, the mass of central region of the density profile was removed and replaced by a point particle with the equivalent removed mass and a suitable softening length (see Table~\ref{tab:sims_inout}). This process removes the steep density profile in that region and can only be applied if a suitable solution for the hydrostatic equilibrium equation is found for a given combination of excised mass and softening length. As a consequence, the softening length needs to be large enough to avoid steep density gradients but small enough that the excised core does not replace more mass than needed from the central region, which would decrease the resolution in that zone. We chose the same softening length value for the companion to avoid adding another parameter to the simulation. It is worth noting that there is no relation between stellar core radii in the original \mesa\ profiles and the softening lengths. Specifically, the stellar cores in the 1.7 and 3.7~\Msun\ models have radii of 0.01 and 0.02~\Rsun, respectively. After the core excision procedure, the modified profile was numerically relaxed following the technique implemented by \cite{Lau2022a} and the results are shown in Figure~\ref{fig:rho-R}. } 

Two equations of state (EoS) were used: an ideal gas EoS and a tabulated EoS that includes the effect of recombination energy. The latter was implemented as in the \mesa\ code and so we often refer to it as the \mesa\ EoS (for details of the implementation see \citealt{Reichardt2020}). In general, models with an ideal gas EoS are more stable, since the internal energy of the gas is smaller than those using a tabulated EoS. Moreover, the 3.7~\Msun\ models are more unstable than the 1.7~\Msun\ ones, likely due to the structure of the star near the core. A larger softening length for the point mass particle increases stability, but reduces the size of the central volume where the interaction is well reproduced. \secondreview{As a precaution, we performed a series of tests on the numerical stability of the 3.7\Msun\ model. These are presented in Appendix~\ref{app:stellar_relaxation}. Notably, we have confirmed that the radial expansion of the 3D model is similar to the original \mesa\ model (Figure~\ref{fig:PD_rad}) and that the envelope is bound to the system in the absence of the companion (Figure~\ref{fig:vel_hist}).}    

\review{Simulations with the 1.7~\Msun\ donor star were started at the time of the Roche lobe overflow in a circular orbit with the companion; the only exception was the model 2-idk15. For that particular simulation, we took a snapshot of model 2-id at $t=7.58$ yr and continued the evolution in a separate model with $\kappa_{\rm max}~=~15$~cm$^2$~g$^{-1}$ (Table~\ref{tab:sims_inout}).}

\review{Simulations with the 3.7~\Msun\ donor star were set with a Roche lobe slightly larger than the stellar radius in a circular orbit with the companion. In particular, the Roche radius $R_{\rm L}=343$~\Rsun\ for the 3.7~\Msun\ stars (Figure~\ref{fig:4M_thermal-pulse}). We used this $R_{\rm L}$ value to create the conditions in which the expansion of the donor star during the pulse is the main cause for the Roche lobe overflow. Furthermore, we avoided selecting a Roche radius too close to previous expansions of the model with similar radii, such as the peak of the previous pulse or the last plateau phase, around a thousand years before the selected moment. Roche radii were calculated using the Eggleton approximation \citep{Eggleton1983}:}   

\begin{equation}
    R_{\rm L}=\frac{0.49q^{2/3}}{0.6q^{2/3}+\ln{\left(1+q^{1/3}\right)}}.
    \label{eq:Roche_Eggleton}
\end{equation}

\review{The interval of the thermal pulse in which the overfilling of the Roche lobe can happen is indicated with a dashed black line in Figure~\ref{fig:4M_thermal-pulse}, and has a duration of 128 years.}

Analysis of the stability of the 1.7~\Msun\  models can be found in \citet{GonzalezBolivar2022}. The stability analysis of the 3.7~\Msun\ model, is described in Appendix~\ref{app:stellar_relaxation}. 

\subsection{Dust driving implementation}
\label{ssec:dust_driving}

\begin{figure}
   \centering
         \includegraphics[width=0.43\textwidth]{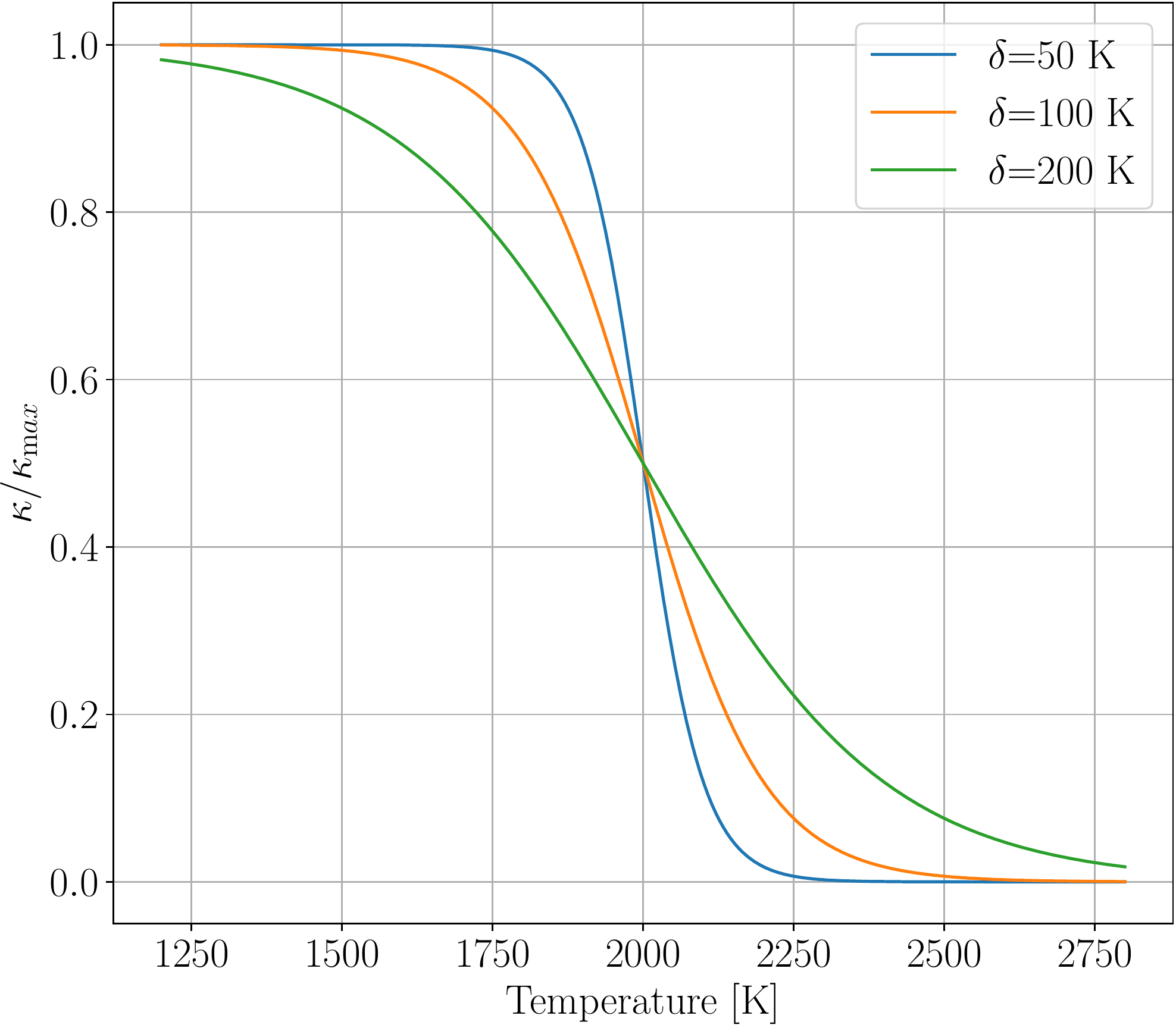}
         \caption{\secondreview{Functional form of the opacity following Bowen prescription for $T_{\rm cond}=2000$. We normalise the opacity and plot three curve for $\delta=50,100$ and 200K.}}
    \label{fig:kapT}
\end{figure}

To simulate the formation of dust, we use the analytical prescription devised by \cite{Bowen1988} for the dust opacity, $\kappa_{\rm D}$, which only depends on the equilibrium temperature, $T_{\rm eq}$ according to
\begin{equation}
    \kappa_{\rm D} = \frac{\kappa_{\rm max}}{1+e^{ (T_{\rm eq}-T_{\rm cond})/\delta }};
    \label{eq:kappa(T)}
\end{equation}
where $\kappa_{\rm max}$ is the maximum dust opacity, $T_{\rm cond}$ is the condensation temperature of the dust, and $\delta$ is the condensation temperature range; these three values are selected and kept constant during each simulation. We use $\kappa_{\rm max}=5$~cm$^2~g^{-1}$ \citep{Bowen1988}, $T_{\rm cond}=2000$~K (this is larger than the canonically utilised 1500~K, \citealt{Hofner2007,Hofner2003,Cristallo2021} and was selected to maximise the effect of dust) and $\delta=200$~K \secondreview{(this means that condensation happens between 1200~K and 2800~K, or $T_{\rm cond} \pm 4\delta$)}. \review{This $\delta$ value allows the opacity function to change smoothly in regions with temperatures around $T_{\rm cond}$. Later on we modify these values to assess their impact on the results, selecting $\kappa_{\rm max}$=15~cm$^2~g^{-1}$ (this value is similar to the peak values in the detailed calculation of Paper~II), $T_{\rm cond}=1500$~K and $\delta=50$~K. The functional form of this prescription, as well as the effect that $\delta$ has in it, is shown in Figure~\ref{fig:kapT}. The shape of the curve is the same for any $T_{\rm cond}$ value. While we do not employ a value of $\delta=100$~K in the simulations, we plotted this for illustrative purposes.}

In Table~\ref{tab:sims_inout} we list the input parameters of all models. The models are named based on the primary mass at the ZAMS (2 or 4), on the equation of state used ("i" for ideal or "m" for \mesa, tabulated EoS) and whether or not dust driving is taken into account (a "d" is added when dust-driving is included). For the simulations including dust driving, we carried out simulations with $\kappa_{\rm max}=5$ and 15 cm$^2$~g$^{-1}$ --- indicated in the name only for values of 15. Finally "dT50" is added to the name when the value of $\delta$ was 50~K instead of 200~K and "T1500"  was added when the condensation temperature was 1500~K instead of 2000~K.

In the absence of a proper treatment of radiative transfer in the simulation, we assume that the radiative equilibrium dust temperature is the same as  the gas temperature. 
The effect of the putative dust on the gas acceleration is introduced in the simulations by modifying the momentum equation in {\phant} according to

\begin{equation}
    \dv{\bm{v}}{t} = -\frac{\nabla P}{\rho} + \bm{a}_{\rm gas}(\bm{r},t) + \bm{a}_{\rm sink-gas} + \frac{\kappa_{\rm D} L}{4\pi r^2 c} \frac{\bm{r}}{r},
    \label{eq:dvdt}
\end{equation}

\noindent where $\bm{v}$, $t$, $\rho$ and $\bm{r}$ are the velocities, time, density and position vector from the core of the primary, respectively; $P$, $\bm{a}_{\rm gas}(\bm{r},t)$ and $\bm{a}_{\rm sink-gas}$ are the pressure, gravitational acceleration due to SPH particles (self-gravity) and due to the sink particles (which represent the core of the AGB star and the companion), respectively. The last term of Eq.~\ref{eq:dvdt} is the contribution of the dust-driven acceleration, $\bm{a}_{\rm dust}$, powered by the luminosity of the primary star, $L$ (and $c$ is the speed of light). For our simulations, we set a constant luminosity, which was calculated by \mesa: $5180$~\Lsun\ and $1.19 \times 10^4$~\Lsun\ for the $1.7$ and $3.7$~\Msun, respectively. The radial coordinate $r$ is taken with respect to the primary's core. We do not account for the luminosity of the companion as it would be thousands, to tens of thousand times weaker.

The Bowen formalism is used to calculate approximate values for dust-driven accelerations in late phase AGB stars, such as Mira variables \citep{Fleischer1992,Chen2020,Bermudez2020,Esseldeurs2023} and was recently implemented in \phant\ by \citet{Siess2022}. It has also been used in various hydrodynamic codes, including SPH, to simulate the interaction of an AGB wind with a distant companion \citep[e.g.][]{Chen2017,Chen2020,Aydi2022,Esseldeurs2023}. Moreover, the dynamical and immediately post-dynamical times of the CE interaction mimic both the time and size-scales of the expansion phase in a pulsating Mira \citep[see discussion in][]{Galaviz2017}, making the Bowen approximation reasonable to determine the opacities in the expanding envelope of giants in the early CE phase. 

%% file: results/results.tex
\section{Results}
\label{sec:results}
\begin{table*}
\centering
\begin{tabular}{lccccccccccr}
\hline
Model & $M_1$ & $q=M_2/M_1$ & $r_{\rm soft,1}$ & $r_{\rm soft,2}$ & EoS  & $a_{\rm i}$ & $a_{\rm ave,f}$ & $\tau_{\rm RLOF}$&$e_{\rm f}$ & M$_{\rm ub}$ & $t_{\rm f}$\\
& (\Msun) & & (\Rsun) & (\Rsun) & & (\Rsun) & (\Rsun)& (yr) &  & (\%) \\ \hline
2-i & $1.7$ & $0.35$ & 2 & 2 & ideal & 550 & 21 & 11 & 0.021 & $\sim$20 & 27.3\\  
2-id & $1.7$ & $0.35$ & 2 & 2 & ideal & 550 & 23 & 7.8 & 0.039 & $\sim$50 & 24.0\\
2-idk15 & $1.7$ & $0.35$ & 2 & 2 & ideal & 550 & 26 & 7.8 & 0.053 & $\sim$70 & 25\\ 
2-m & $1.7$ & $0.35$ & 2.5 & 2.5 & \mesa & 550 & 32 & 0.7 & 0.002 & $\sim$95 & 12.6\\ 
2-md & $1.7$ & $0.35$ & 2.5 & 2.5 & \mesa & 550 & 33 & 0.7 & 0.009 & $\sim$90 & 12.6\\
2-mdk15 & $1.7$ & $0.35$ & 2.5 & 2.5 & \mesa & 550 & 40 & 0.7 & 0.009 & $\sim$100 & 12.6 \\ 
2-mdk15dT50 & $1.7$ & $0.35$ & 2.5 & 2.5 & \mesa & 550 & 40 & 0.7 & 0.005 & $\sim$95 & 12.6 \\ 
2-mdk15dT50T1500 & $1.7$ & $0.35$ & 2.5 & 2.5 & \mesa & 550 & 40 & 0.7 & 0.012 & $\sim$100 & 12.6 \\ 
4-i & $3.7$ & $0.16$ & 5 & 5 & ideal & 637 & 8.2 & 7.5 & 0.038 & $\sim$20 & 27.4 \\ 
4-id & $3.7$ & $0.16$ & 5& 5 & ideal & 637 & 7.9 & 7.5 & 0.040 & $\sim$40 & 24.0\\
4-m & $3.7$ & $0.16$ & 8 & 8  & \mesa & 637 & 9.4 & 0.8 & 0.022 & $\sim$95 & 25.2\\   
4-md & $3.7$ & $0.16$ & 8 & 8 & \mesa & 637 & 9.9 & 0.8 & 0.021 & $\sim$100 & 12.6 \\ 
4-mdk15  & $3.7$ & $0.16$ & 8 & 8 & \mesa & 637 & 12 & 0.6 & 0.032 & $\sim$100 & 12.6 \\
4-mdk15dT50  & $3.7$ & $0.16$ & 8 & 8 & \mesa & 637 & 12 & 0.6 & 0.046 & $\sim$100 & 12.6 \\
4-mdk15dT50T1500 & $3.7$ & $0.16$ & 8 & 8 & \mesa & 637 & 10 & 0.6 & 0.032 & $\sim$100 & $12.6 $\\ \hline 

\end{tabular}
\caption{Parameters of the simulated binary stellar systems. The first column is the model name for each simulation (which indicates the ZAMS mass of the giant's model). $M_1$ is the donor's mass (including the point mass particle) of each simulation, $q=M_2/M_1$ is the mass ratio of the binary ($M_2=0.6$~\Msun), $r_{\rm soft,1}$ and $r_{\rm soft,2}$ are the softening lengths of the point mass particles, EoS is the equation of state used ($i$ for ideal, $m$ for \mesa\ or tabulated equation of state); $a_{\rm i}$ is the initial orbital separation, $a_{\rm ave,f}$ is the final average separation (defined as the average of the last periastron and apastron distances of the simulation), $e_{\rm f}$ is the eccentricity at the end of the simulation and $M_{\rm ub}$ is the percentage of unbound envelope \review{at the end of the simulation (except for models 2-i and 2-id whose $M_{\rm ub}$ values were taken at the end time of model 2-idk15 for ease of comparison). Finally, $t_{\rm f}$, is the physical time duration of each simulation.}}
 \label{tab:sims_inout}
\end{table*}


\begin{figure*}
     \centering
     \begin{subfigure}[b]{0.48\textwidth}
         \centering
         \includegraphics[width=\textwidth]{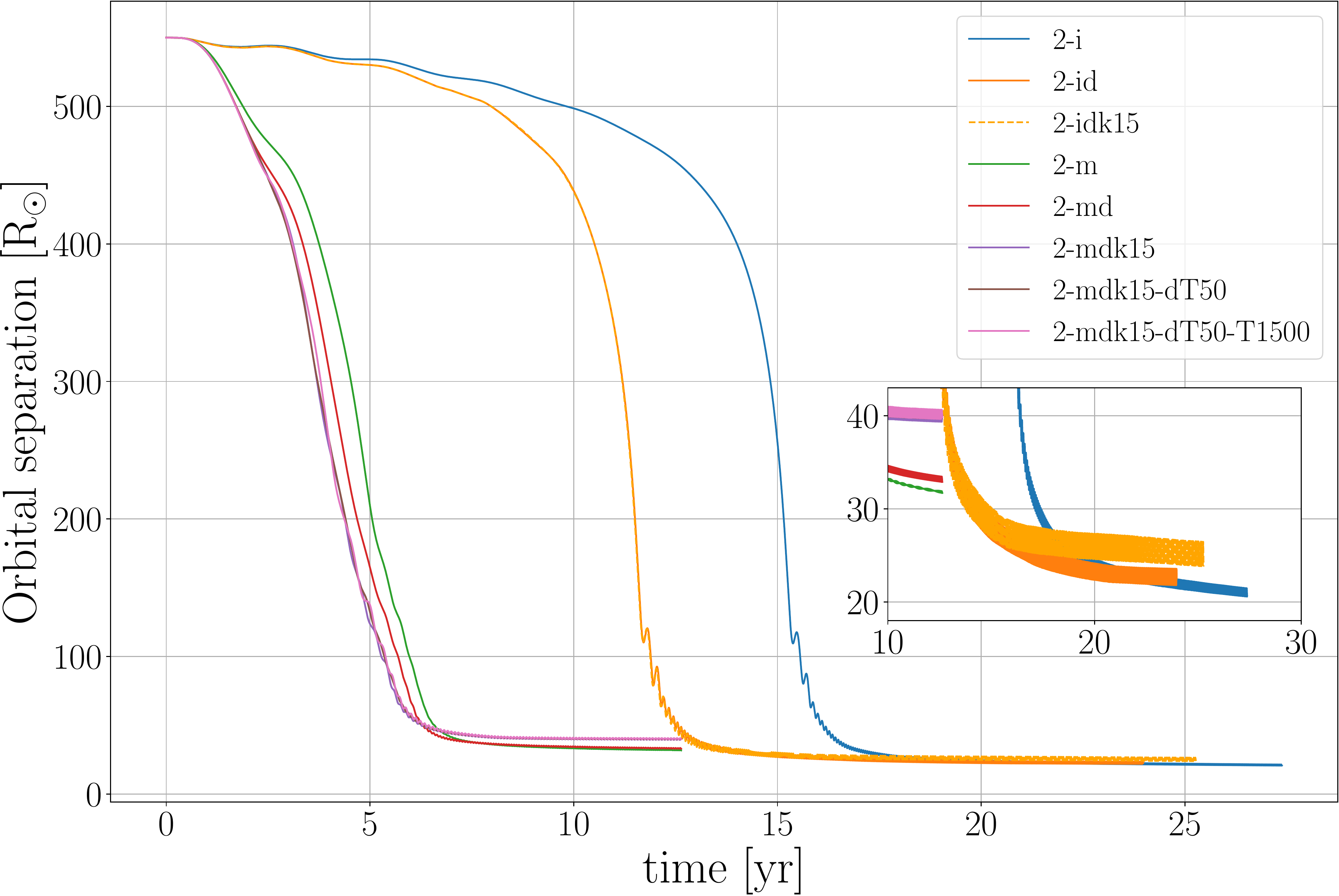}
         \end{subfigure}
     \hfill
     \begin{subfigure}[b]{0.48\textwidth}
         \centering
         \includegraphics[width=\textwidth]{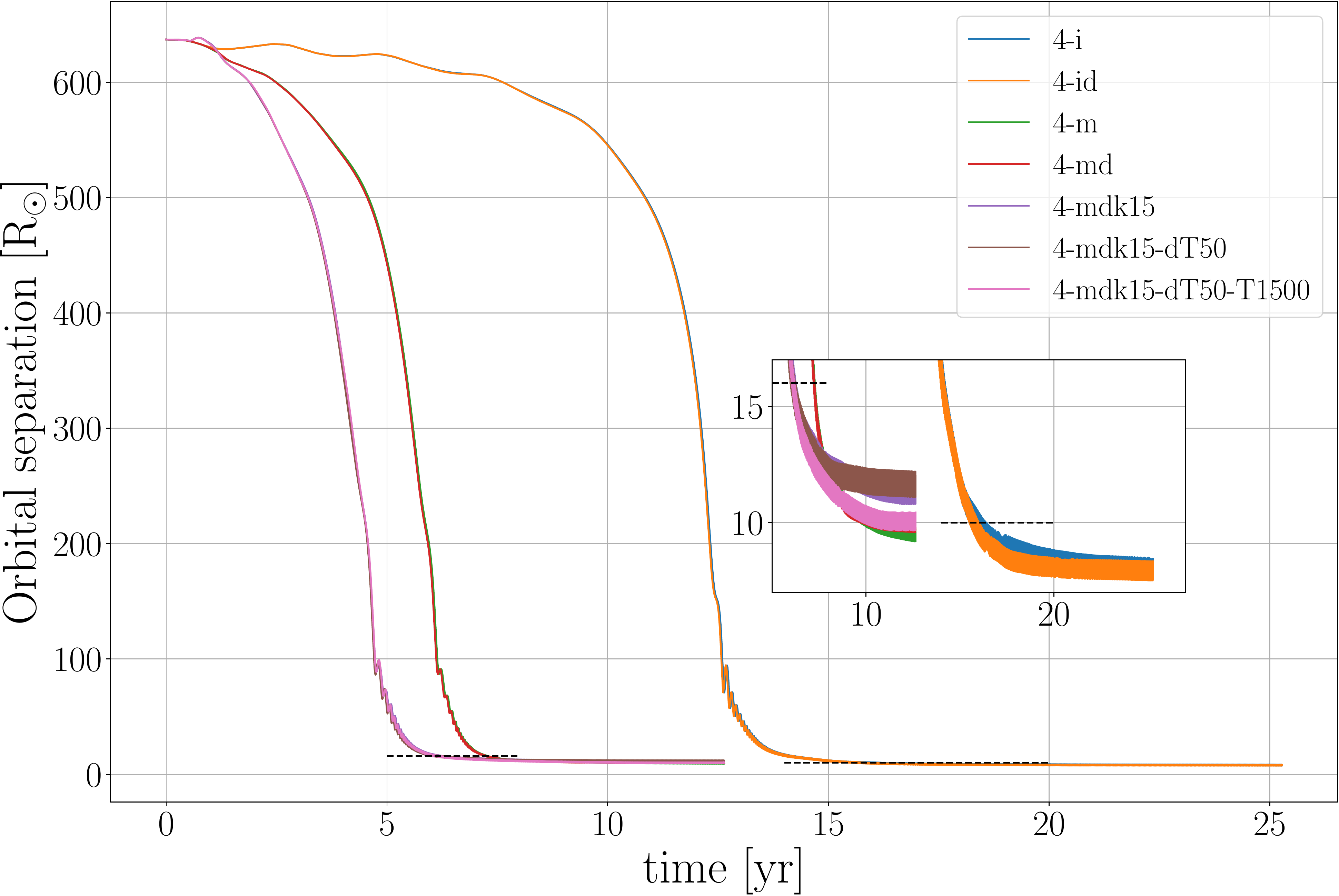}
         \end{subfigure}
     \hfill
     \centering
     \begin{subfigure}[b]{0.48\textwidth}
         \centering
         \includegraphics[width=\textwidth]{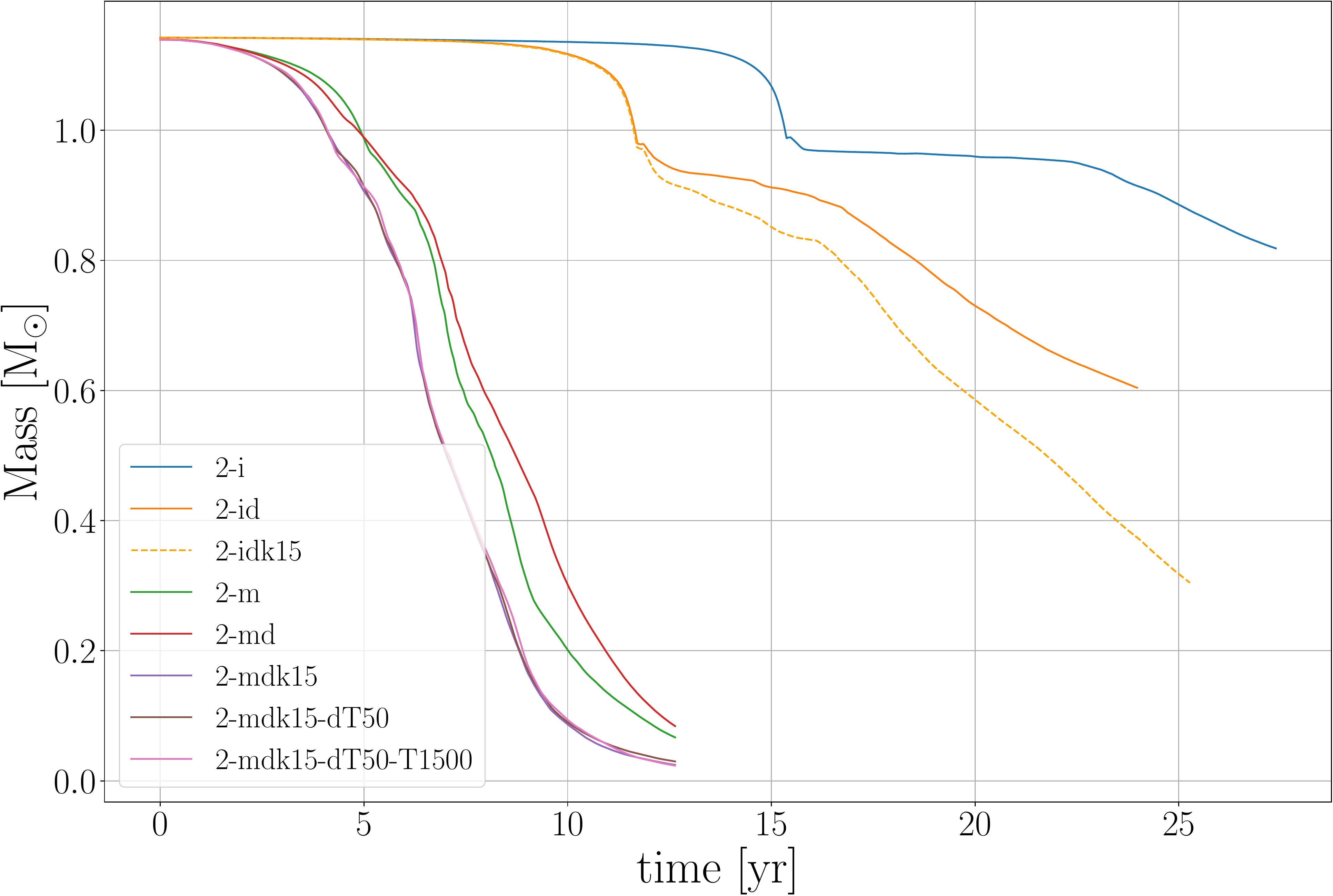}
         \end{subfigure}
     \hfill
     \begin{subfigure}[b]{0.48\textwidth}
         \centering
         \includegraphics[width=\textwidth]{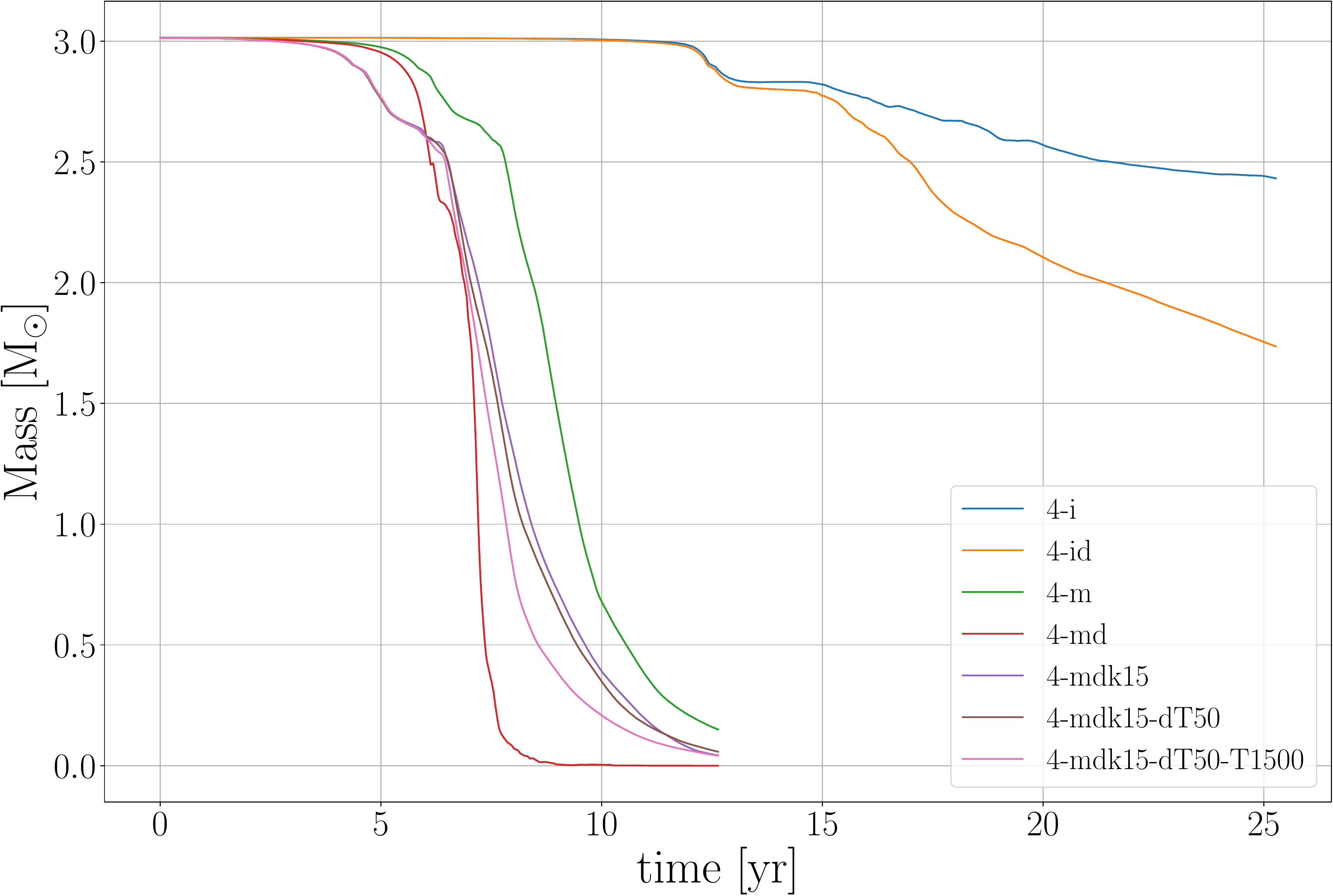}
         \end{subfigure}
     \hfill
     \caption{Orbital evolution (top panels) and envelope bound mass (bottom panels) of all models. Left panel: 1.7~\Msun{} simulations have final separations that range between 20-32 \Rsun. Right panel: the 3.7~\Msun\ models finish with separations between 7-11 \Rsun. The 2-id and 2-idk15 are almost identical and only slightly different at the very end of the simulation. Envelope bound mass for all models. Both 1.7~\Msun\ (left panel) and 3.7~\Msun\ (right panel) models have a dependency on the equation of state implemented in the simulation.}
    \label{fig:boundsep}
\end{figure*}

The input and some output parameters for all simulations are shown in Table~\ref{tab:sims_inout}. The final separation, $a_{\rm ave,f}$, was calculated as the average of the periastron and apastron of the binary at the end of the simulation. The RLOF timescale, $\tau_{\rm ROLF}$, was calculated  using a modified version of the criterion implemented by \cite{Nandez2016} for the plunge-in phase. This timescale is the interval between the start of the simulation\footnote{For simulation 2-idk15, we have used the initial value of the orbital separation of simulation 2-id, because 2-idk15 does not start at RLOF, as explained in Section~\ref{ssec:dust_driving}.} and the moment where the following condition is first satisfied: 
\begin{equation}
    \frac{|\dot{a}P|}{a}<0.1,
    \label{eq:rolf_crit}
\end{equation}
where $a$ is the orbital separation (distance between the point mass particles), $P$ is the period of their Keplerian orbit and $\dot{a}$ is the rate of change of $a$. While $\tau_{\rm ROLF}$ is resolution dependent \citep[e.g.,][]{Reichardt2019}, for equal resolution, numbers in Table \ref{tab:sims_inout}, column 10 indicate that when dust-driving is accounted for, the time of in-spiral is hastened. This is described in more detail in Section~\ref{ssec:orbital_evolution}. The final eccentricity $e_{\rm f}$ was calculated as 
\begin{equation}
    e_\mathrm{f}=\frac{r_{\rm a}-r_{\rm p}}{r_{\rm a}+r_{\rm p}},
    \label{eeq:cc}
\end{equation}
where $r_{\rm a}$ and $r_{\rm p}$ are the apastron and periastron values used for $a_{\rm ave,f}$. \review{For $r_{\rm a}$ and $r_{\rm p}$ we used the last local maximum in minimum values of the orbital separation values in each simulation, respectively.} The amount of unbound mass at the end of the simulation, $M_{\rm ub}$ is calculated using a criterion where the sum of its potential, kinetic and internal energies is larger than zero.

\subsection{Orbital evolution and unbound mass}
\label{ssec:orbital_evolution}  

Before comparing the differences in the models with and without dust-driving, we describe the salient features of the 3.7~\Msun\ model, which, contrary to the 1.7~\Msun\ model, were not presented elsewhere.

\subsubsection{Common envelope interaction between a 3.7~\Msun\ AGB star and a 0.6~\Msun\ companion (no dust-driving)}


The RLOF  phase and the in-spiral of the 3.7~\Msun\ (non-dusty) models (4-i and 4-m) are qualitatively very similar to the respective models carried out with a 1.7~\Msun\ donor star. The inclusion of recombination energy via the tabulated EoS \citep{GonzalezBolivar2022} cut in half the RLOF timescale compared to the same model run with an ideal gas EoS and but \review{a slight increment of unbound gas, from $95$~per cent to 100 per cent} of the envelope depending on the combination of parameters used in the simulation. This can be seen in both Table~\ref{tab:sims_inout} and Figure~\ref{fig:boundsep}. Once again, we do not place special emphasis on the actual values of the unbound mass except as a comparative measure. 

We cannot comment on the final separations of the 3.7~\Msun\ models because the values are of the order of the core softening length. Within this volume, gravity is softened and the orbital separation evolution is not reliable. We can therefore only state that the final separation for this model is smaller than $\sim 5-10$~\Rsun, as expected for such a small $q=M_2/M_1$ value. 

\review{As mentioned before, the core radius and the softening length of the primary point mass particles are not the same. However, by considering the original core radius of the MESA stellar profile (before mapping into \phant) and the Roche potential of the point mass particles, we inferred that, because of its small radius (0.01\Rsun), the primary core would present no deformation \secondreview{to} the companion at the orbital distance observed at the end of the simulation. Regarding the companion, the possibility of a tidal deformation at the end of the simulation would depend on the type of star that we assume: main sequence or white dwarf. Using the final separations as reference, in neither case the companion for the 1.7~\Msun\ would show tidal deformation due to the size of the Hill sphere. In the binary with the 3.7~\Msun\ donor, only a 0.6\Msun\ main sequence star would present a tidal deformation. 
}

The resolution-dependent mass-unbinding phase, discussed previously by \cite{Reichardt2019}, \citet{Lau2022a} and \citet{GonzalezBolivar2022}, is difficult to diagnose in the absence of a higher resolution simulation. For the ideal gas simulations we have noted previously that a downturn in the unbound mass after the initial mechanical unbinding can be a symptom of artificial unbinding. This happens at $\sim$15 years in 4-i and 4-id (see blue and orange curves in Figure~\ref{fig:boundsep}, bottom right panel. This is why we quote the unbound mass at that time, as a lower limit for these simulations. 

For the 4-m simulation, with a great deal more unbinding it is hard to notice. By visualising the number of unbound particles that originate very close to the core and which propagate in specific directions where density is low \citep[such as evacuated cones near the polar directions, see e.g. Figure 10 of][]{GonzalezBolivar2022} we suspect that artificial unbinding starts at $\sim$9.5~yr. However, the typical number of particles affected by this phase of gas unbinding represents only a few percent of the total and therefore may not be contributing significantly to the amount of unbound mass. Even with the most conservative approach to the artificial unbinding problem, the mass unbound is 95~per cent of the envelope, indicating that full unbinding may well be a simple matter of time.

\subsubsection{The effect of dust-acceleration on the in-spiral and on the unbound mass of the 1.7 and 3.7~\Msun\ models}

\begin{figure*}
     \centering
     \includegraphics[width=\textwidth]{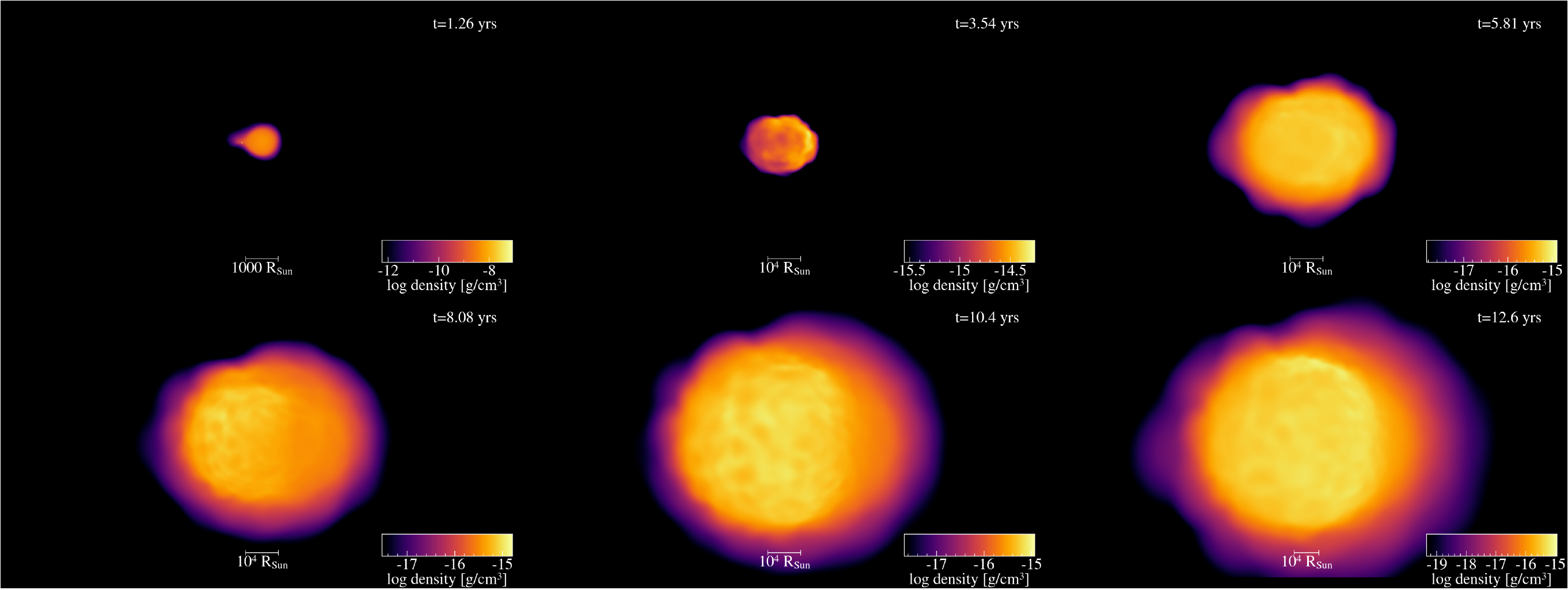}
     \caption{Surface rendering of the density of material at the last scattering surface in model 4-mdk15dT50T1500 using the Bowen opacity at several times during the common envelope. From left to right and top to bottom, each panel is taken at 1.26, 3.54, 5.81, 8.08, 10.4, and 12.6 yr. This image, along with all other renderings of the simulation in this paper, was created with {\sc splash} \citep{Price2007}. Unlike previous work that use a constant opacity value for rendering \citep{Reichardt2019,Lau2022a}, these surfaces use the Bowen $\kappa$. The density is measured at $\tau=1$.}
     \label{fig:rho_kappa_3D}
\end{figure*}

Using the Bowen opacity we can track the expansion of the photosphere in the dusty simulations. For instance, in Figure~\ref{fig:rho_kappa_3D}, we show a set of surface renderings of the 3.7~\Msun\ model. We visualise the density at the photosphere ($\tau \sim 1$). Clearly the object increases in size enormously when dust is accounted for: at time 5.8 year the photosphere has an approximate radius of $\sim$120~au and a density of $10^{-16}$~g~cm$^{-3}$, while at 12.6 years, it is $\sim$240~au with a density of $10^{-14}$~g~cm$^{-3}$. Comparing it to the 4-m simulation at 12.6~yr, using a low opacity of $10^{-4}$~cm$^2$~g$^{-1}$ for neutral gas with $T \lesssim 6000$~K, and an electron scattering opacity of 0.35~cm$^2$~g$^{-1}$ for hotter ionised gas, we obtain a value of $\sim$30~au, noting that the photosphere resides at the ionisation/recombination front. Similarly for the 1.7~\Msun\ model with no dusty (2-m) the photospheric radius is $\sim$20~au at 12.6~yrs, while with dust (2-md) it is $\sim$170~au at the same time.

Other noticeable effects of including dust-driven accelerations using the Bowen approximation is the tendency to a shortening of the RLOF phase (for some of the models more than for others), a somewhat faster and slightly larger amount of unbound mass, and a slightly larger final separation. Below we give details for each model.

For the 1.7~\Msun\ simulations with an ideal gas EoS the RLOF timescale ($\tau_\mathrm{RLOF}$) is reduced by $30\%$ when the dust-driven acceleration is accounted for (Table~\ref{tab:sims_inout}, column 9) independently of the value of the maximum opacity. This effect is not as pronounced for the case of the tabulated EoS. Interestingly, the 3.7~\Msun, ideal gas EoS simulation, with dust-driven accelerations displays an identical inspiral to the inspiral with no dust-driving, possibly because the lower value of $q$ does not instigate early expansion.
For the equivalent tabulated EoS simulations, dust-driving  results in a shorter value for $\tau_\mathrm{RLOF}$, but only for larger dust opacities (4-mdk15 models). 

\begin{figure}
     \centering
     \includegraphics[width=0.47\textwidth]{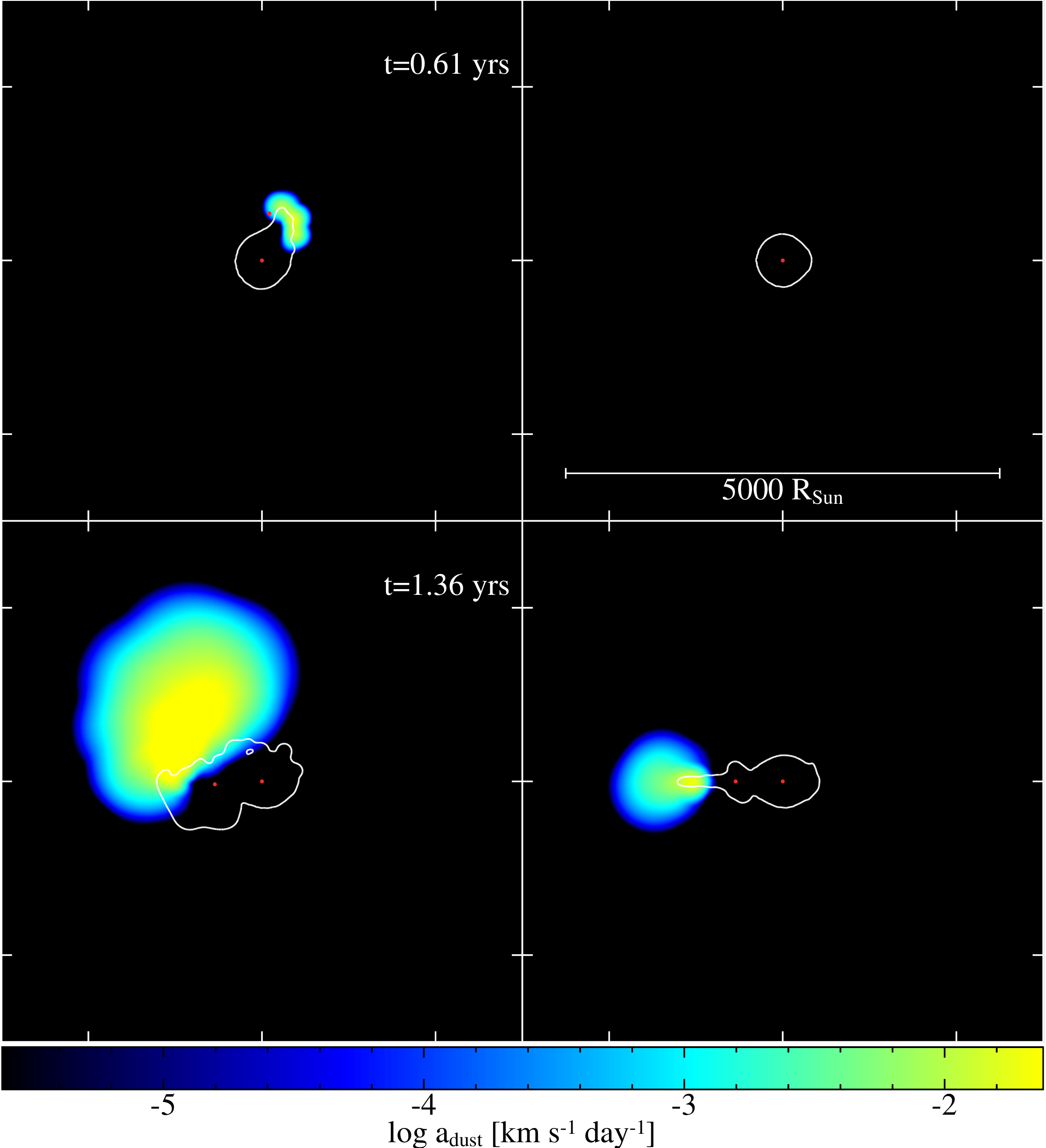}
     \caption{Cross-sections showing the gas acceleration due to dust-driving ($a_\mathrm{dust}$) for the 2-id simulation (Table~\ref{tab:sims_inout}) at early times. Left panels show the orbital (x-y) plane of the binary (shown at z=0) and right panels the perpendicular (z-x) plane (shown at y=0). The white contour line indicates $T=2000$~K. Red dots represent the point mass particles.} 
     \label{fig:early_dvdt}
\end{figure}

The shortening of the RLOF time is an effect of the increase in opacity at the surface of the star due to adiabatic gas cooling at early stages in the simulation. Figure~\ref{fig:early_dvdt} shows the magnitude of the additional dust-driven acceleration 
for model 2-id. The gas that is accelerated by the companion point mass particle at $t\approx0.6$ yrs is ejected from the system with higher velocities compared to the non dust-driven model. This small additional velocity increases the pressure gradient  so that more SPH particles are accelerated, which results in a faster mass transfer. This effect is much less pronounced in the 3.7~\Msun\ simulations.

The 1.7~\Msun\ models with dust-driving result in slightly larger final separations, in particular those that use a tabulated EoS. Increasing $\kappa_\mathrm{max}$ also increase noticeably the final separation ($\sim$40~\Rsun, compared to $\sim$33~\Rsun). For the 3.7~\Msun\ models we cannot comment because the final separations are in all case smaller than the core softening length.

\review{In general, dust-driven acceleration increases the final eccentricity of the orbit (Table~\ref{tab:sims_inout}, column 10). For all 1.7~\Msun\ models, values of $e_{\rm f}$ are larger for models with dust. For the 3.7~\Msun, this is mostly true as well, but for some cases the values are slightly smaller. For example, 4-m has a larger final eccentricity than model 4-md, but both values are smaller than the rest of the dusty models with this donor star using recombination energy. This suggests that the faster SPH particles interact with the companion, taking more angular momentum of the point mass particle during the plunge-in. This behaviour is further increased for larger values of $\kappa_{\rm max}$, where the dust-driven acceleration is ~3 times larger and the eccentricity is larger compared to the non dusty and dusty models with $\kappa_{\rm max}=5$~cm$^2$~g$^{-1}$.} 

For all the simulations computed with the tabulated EoS, the behaviour is more complex and not consistent between the 1.7 and 3.7~\Msun\ simulations. Overall there are only small differences between simulations with and without dust-driving. The envelope expansion is generally promoted by an initially larger and slightly less stable stellar structure \citep[see discussion in][]{GonzalezBolivar2022} and, later, by the release of recombination energy. Here the additional acceleration due to dust-driving is weak and blurred by the impact of the recombination energy on the dynamics.

Overall the slightly larger amount of unbound mass in all models with dust-driving can be ascribed to timing (mass is unbound earlier). Resolution-dependent mass unbinding may be taking place in all simulations with a tabulated EoS, towards the end. We have carried out an identical analysis to that of \citet{GonzalezBolivar2022} and listed in Table~\ref{tab:sims_inout} the times at which cones of unbound particles appear near the central binary. It is at these times that we also list the amount of unbound mass. This is a very conservative approach because the fact that mass is unbound near the central binary in low density cones can be a physical phenomenon. Also, the amount of mass unbound in these locations is very small, equivalent to less than 5~per cent of the total mass of the envelope, lending credit to the fact that the unbound mass curves in Figure~\ref{fig:boundsep} (bottom panels) are reasonably accurate to the end of the simulations. As we have stated above, there is only one way to determine whether mass unbinding is resolution dependent, and that is to carry out a higher resolution simulations ($\sim$10 million SPH particles), something that is not possible at this time.

\begin{figure*}
     \begin{subfigure}[b]{0.48\textwidth}
         \centering
         \includegraphics[width=\textwidth]{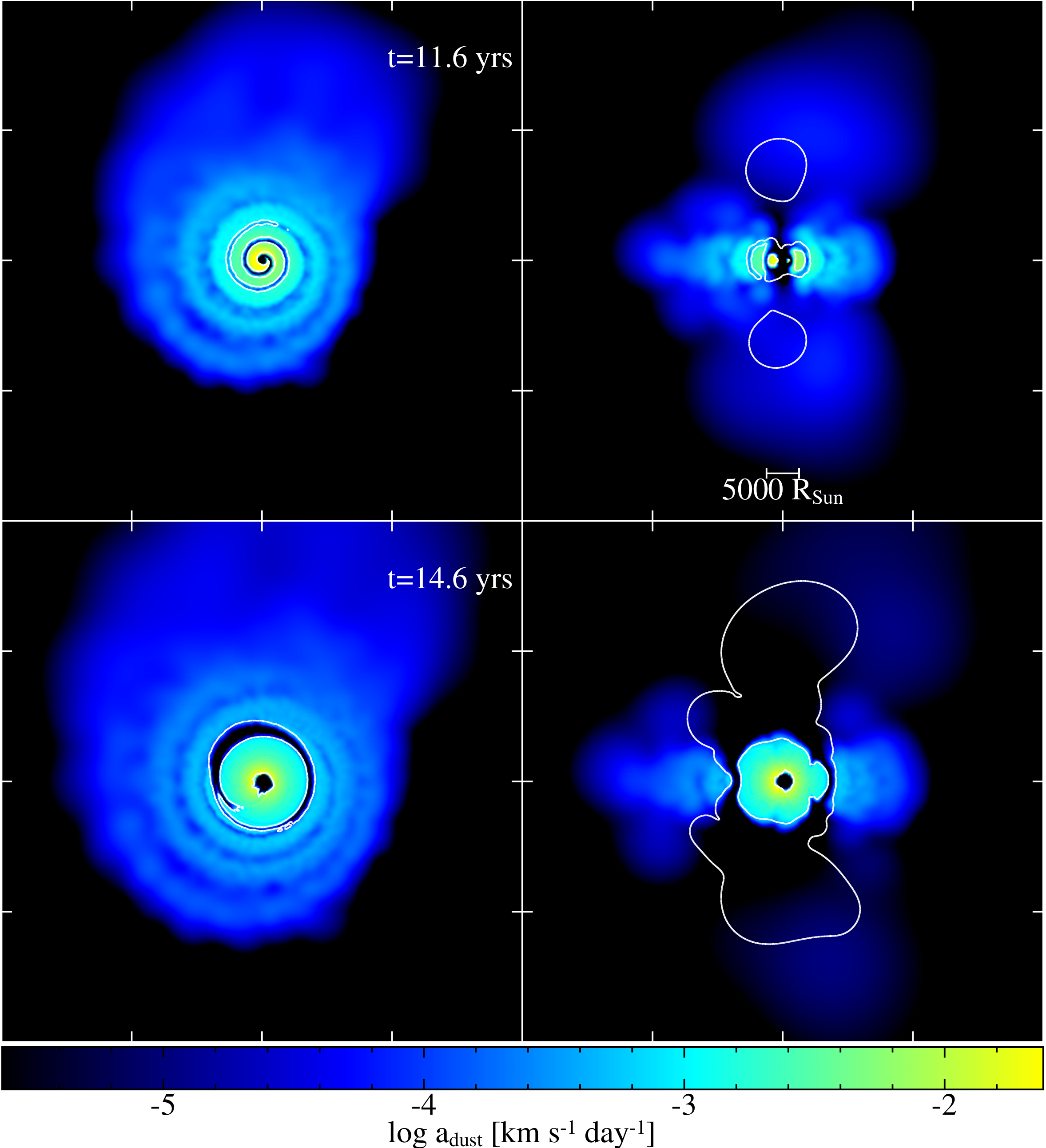}
         \end{subfigure}
     \hfill
     \begin{subfigure}[b]{0.48\textwidth}
         \centering
         \includegraphics[width=\textwidth]{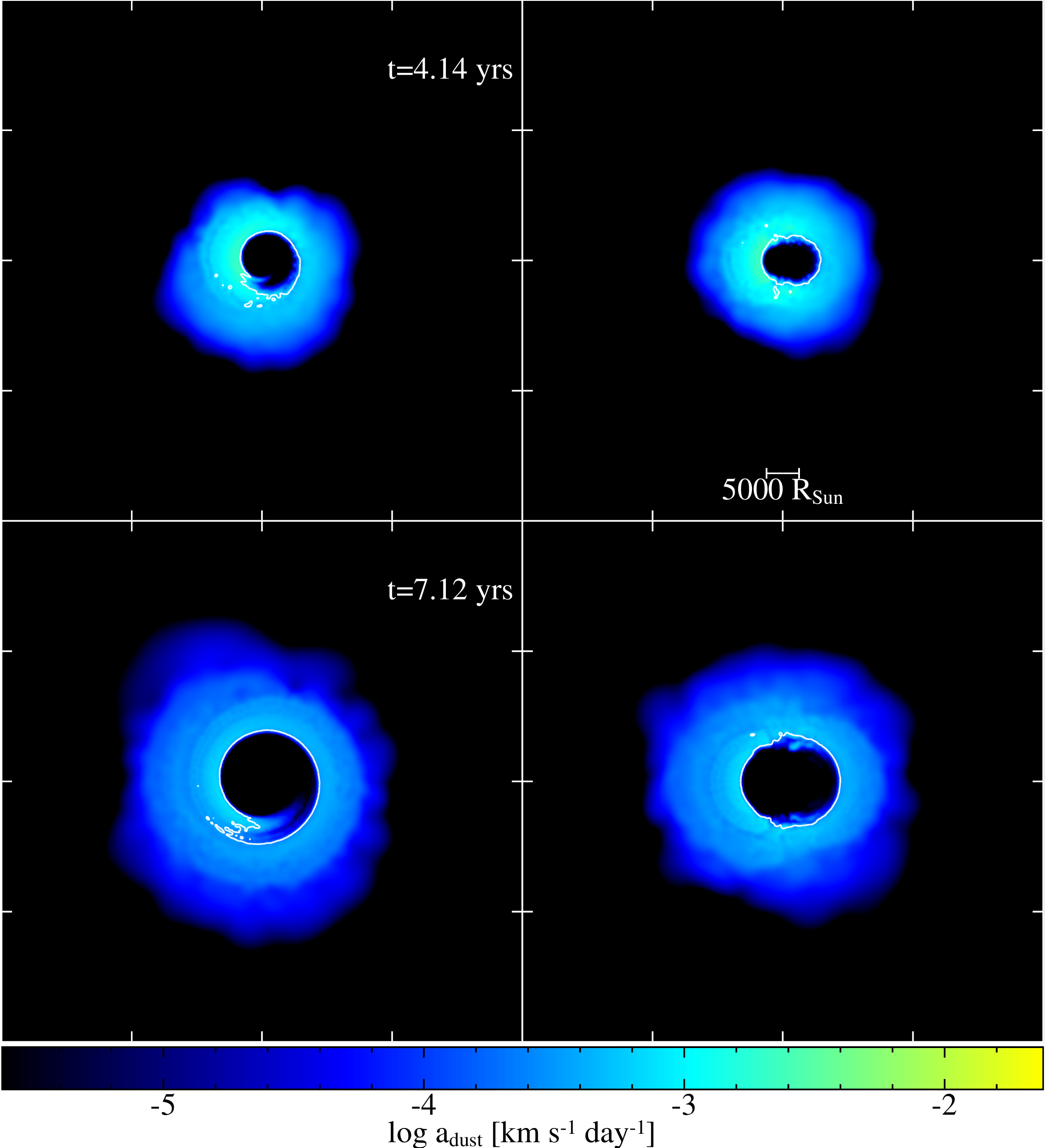}
         \end{subfigure}
     \hfill
     \caption{Cross sections of the dust-driven acceleration ($a_\mathrm{dust}$) for 2-id (left) and 2-md (right) models. First and third columns show the orbital plane (x-y) at $z=0$. Second and fourth columns show the perpendicular plane to the orbit (x-z) at $y=0$. The contours (white lines) are at T=2000K.}
    \label{fig:arho2M}
\end{figure*}

\begin{figure*}
     \centering
     \begin{subfigure}[b]{0.48\textwidth}
         \centering
         \includegraphics[width=\textwidth]{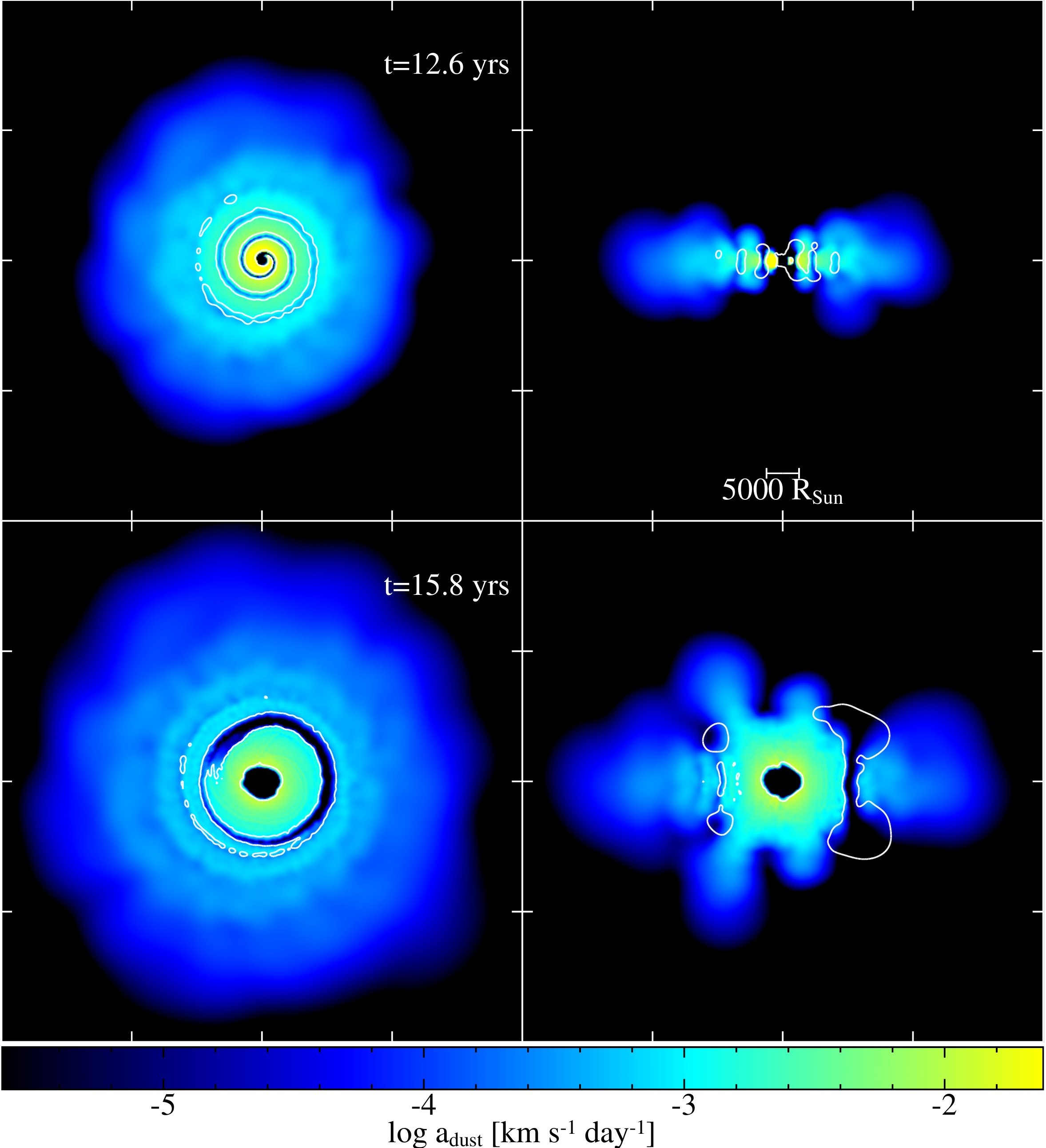}
         \end{subfigure}
     \hfill
     \begin{subfigure}[b]{0.48\textwidth}
         \centering
         \includegraphics[width=\textwidth]{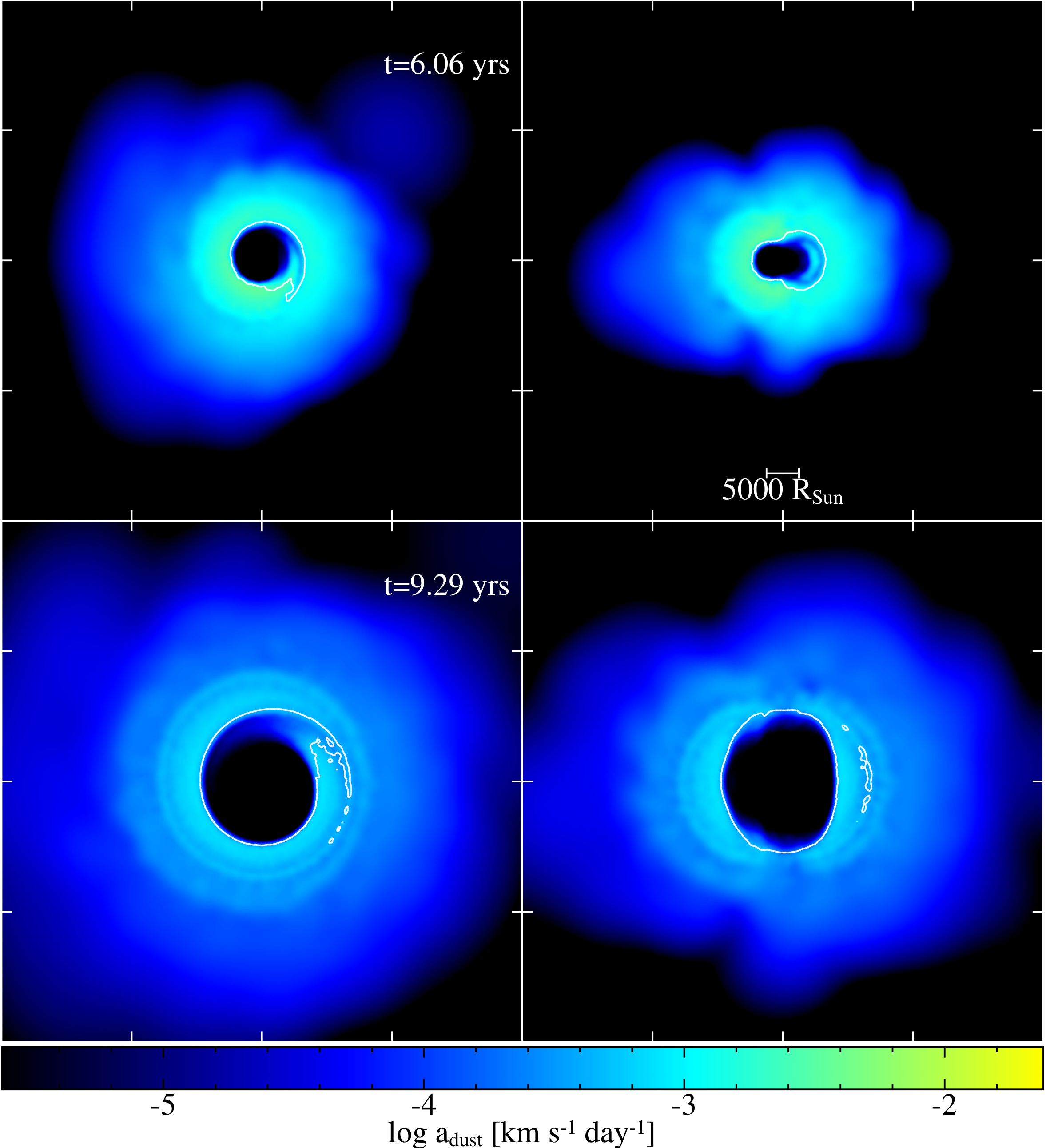}
         \end{subfigure}
     \hfill
     \caption{Dust-driven acceleration  ($a_\mathrm{dust}$) cross sections for the 4-id (left) and 4-md (right) models. First and third columns show the orbital plane (x-y) at $z=0$. Second and fourth columns show the perpendicular plane to the orbit (x-z) at $y=0$. The contours (white lines) are at T=2000K.}
    \label{fig:arho4M}
\end{figure*}

Figures~\ref{fig:arho2M} and \ref{fig:arho4M} show the dust-driven acceleration of the gas for models 2-id, 2-md, 4-id and 4-md. The times selected correspond to the moment of the steepest in-spiral and the end of the in-spiral as defined by the moment after the in-spiral when $\dot{a}P/a 
 > 0.1$. The yellow-green region corresponds to a dust-driven acceleration of $\sim 10^{-2}$~km\ s$^{-1}~\rm{day}^{-1}$ which, if constant, increases the radius by $\sim 6$~\Rsun\ every 100 days, compared to gas without dust-driven acceleration (assuming all the vector terms in the right hand side of Eq.~\ref{eq:dvdt} are aligned). Ideal gas models have larger dust-driven accelerations near the central region. Interestingly, in the equatorial plane, the gas contained in the spiral has a higher temperature than $T_{\rm cond}$ (white contour), but as it moves outwards, it quickly expands and cools down  effectively increasing its dust-driven acceleration compared to the gas inside the spiral. 

On the other hand, simulations with recombination energy (right panels, Figures~\ref{fig:arho2M} and \ref{fig:arho4M}) have overall smaller dust-driven accelerations and lack spiral arm-like structures. This is because accelerations due to the evenly distributed release of recombination energy smooth out the density field. As was already noticed by \citet{GonzalezBolivar2022}, models with recombination energy display more symmetry explaining the approximately spherical distribution of the dust-driven accelerations. \review{}

\subsection{Mass and velocity of the dust-driven gas}
\label{sec:mvgas}
\begin{figure*}
     \centering
     \begin{subfigure}[b]{0.48\textwidth}
         \centering
         \includegraphics[width=\textwidth]{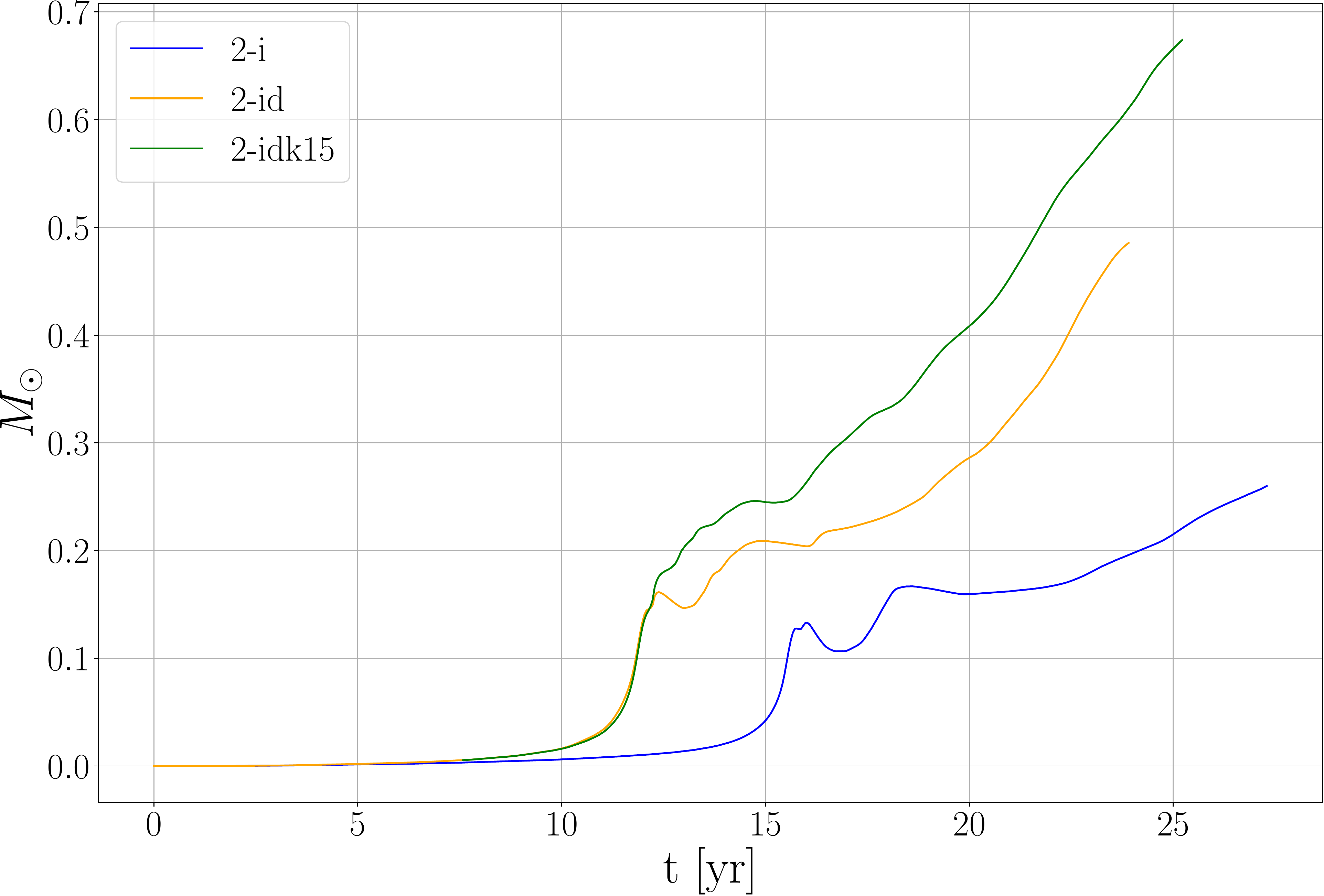}
         \end{subfigure}
     \hfill
     \begin{subfigure}[b]{0.48\textwidth}
         \centering
         \includegraphics[width=\textwidth]{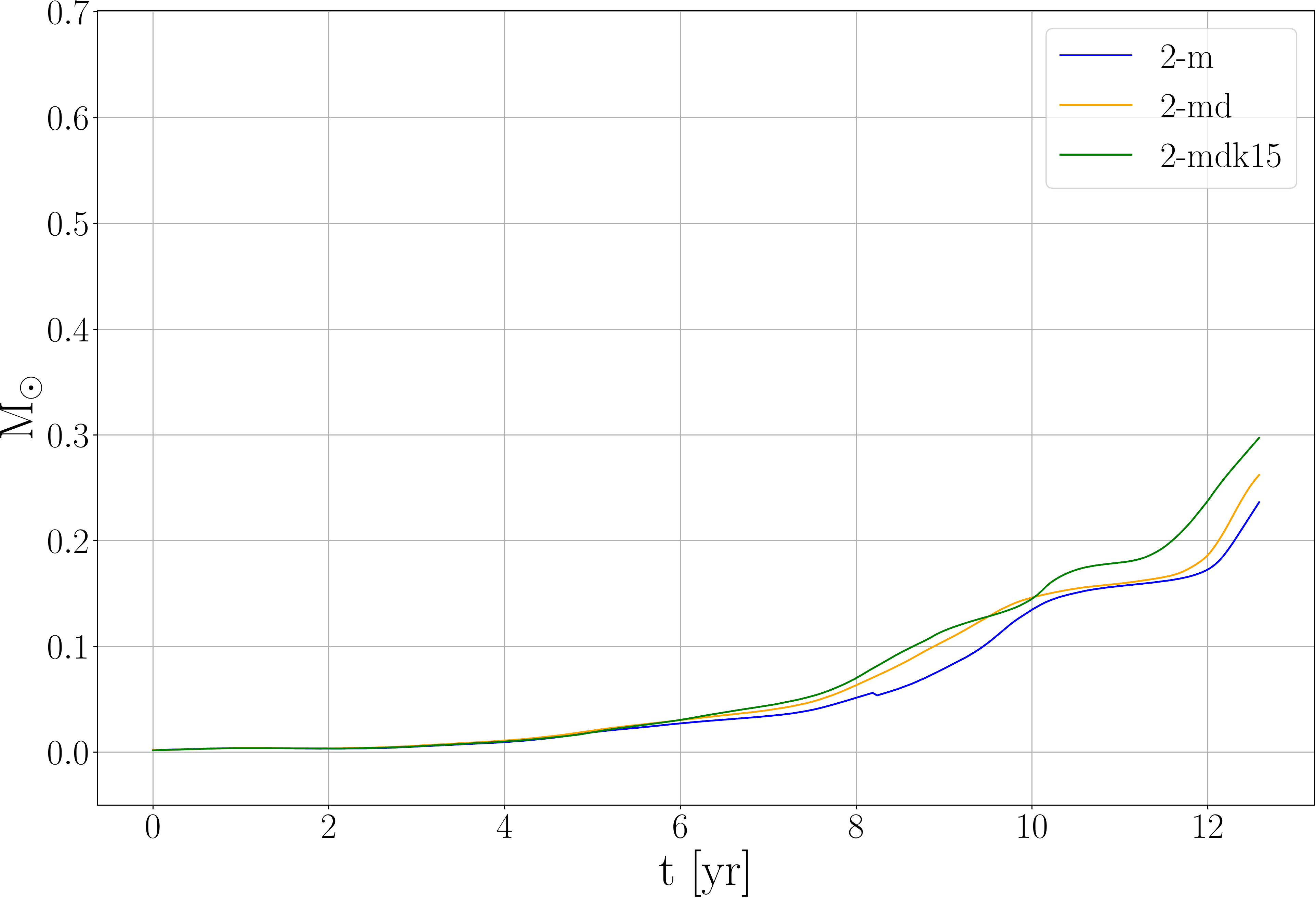}
         \end{subfigure}
     \hfill
     \centering
     \begin{subfigure}[b]{0.48\textwidth}
         \centering
         \includegraphics[width=\textwidth]{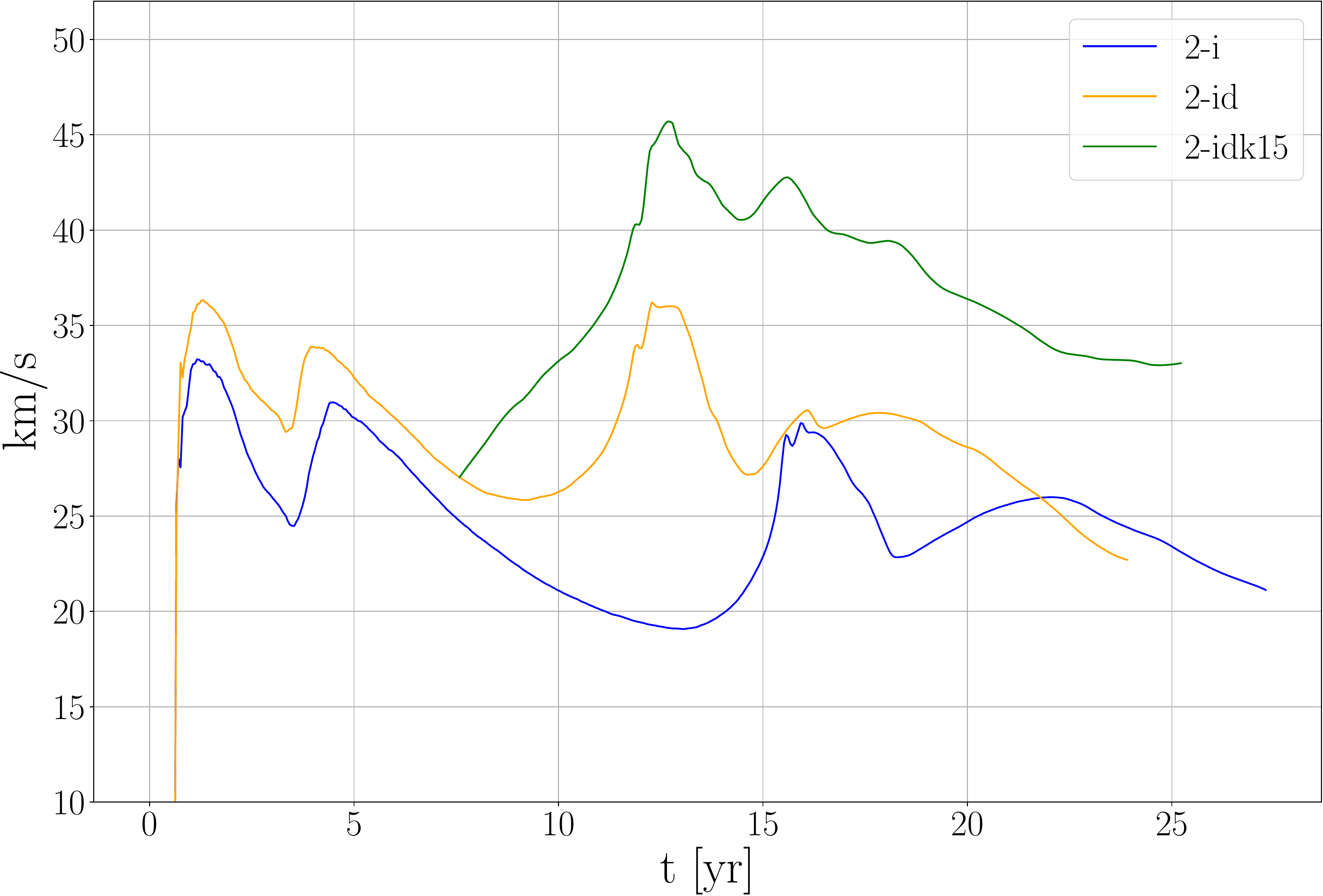}
         \end{subfigure}
     \hfill
     \begin{subfigure}[b]{0.48\textwidth}
         \centering
         \includegraphics[width=\textwidth]{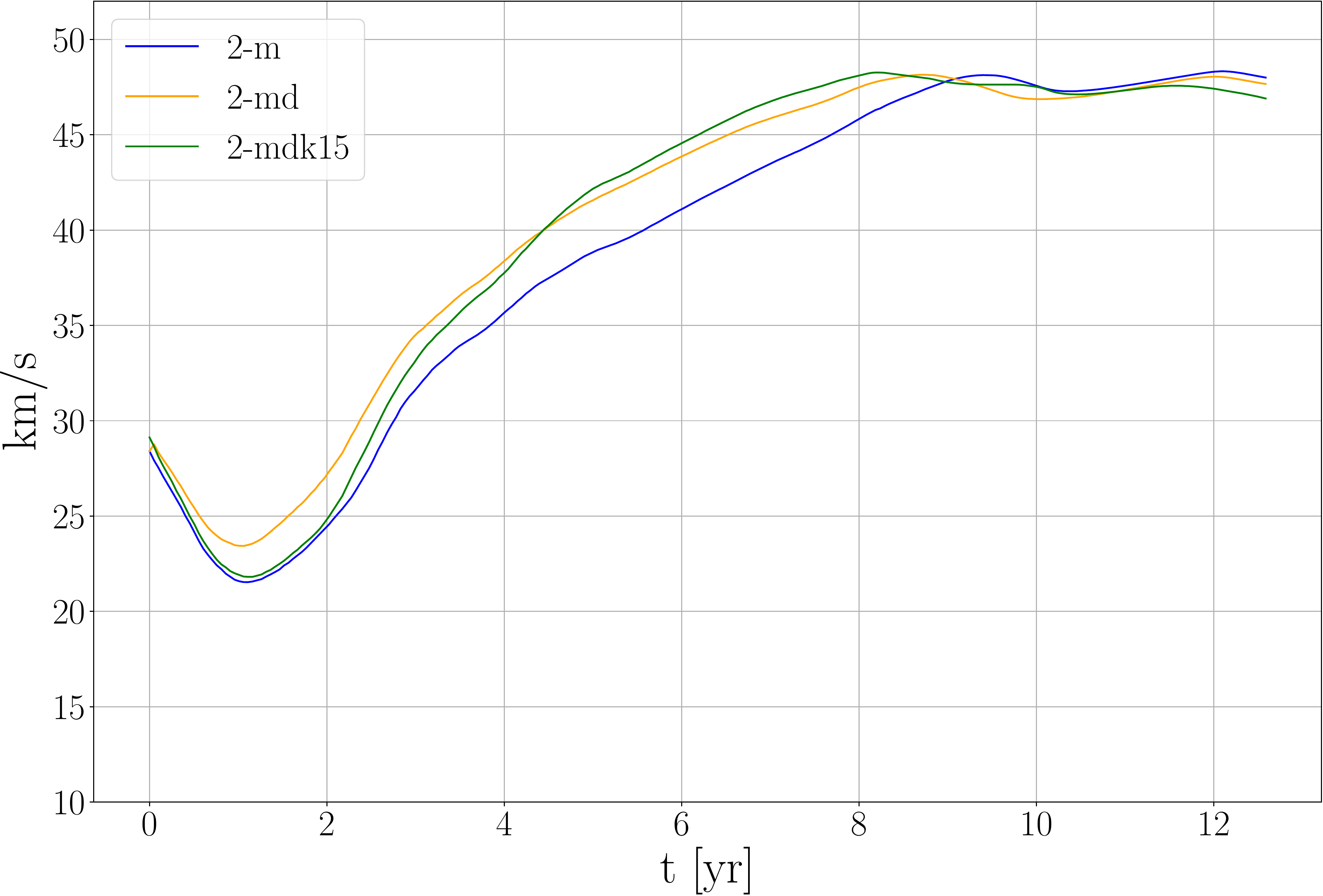}
         \end{subfigure}
     \hfill
     \caption{Top row: The total mass of SPH particles with a temperature lower than 2000~K for the 2~\Msun\ simulation. Bottom panels: average speed of the gas with T$<2000$~K  for the same simulations.}
    \label{fig:mv2M}
\end{figure*}

\begin{figure*}
     \centering
     \begin{subfigure}[b]{0.48\textwidth}
         \centering
         \includegraphics[width=\textwidth]{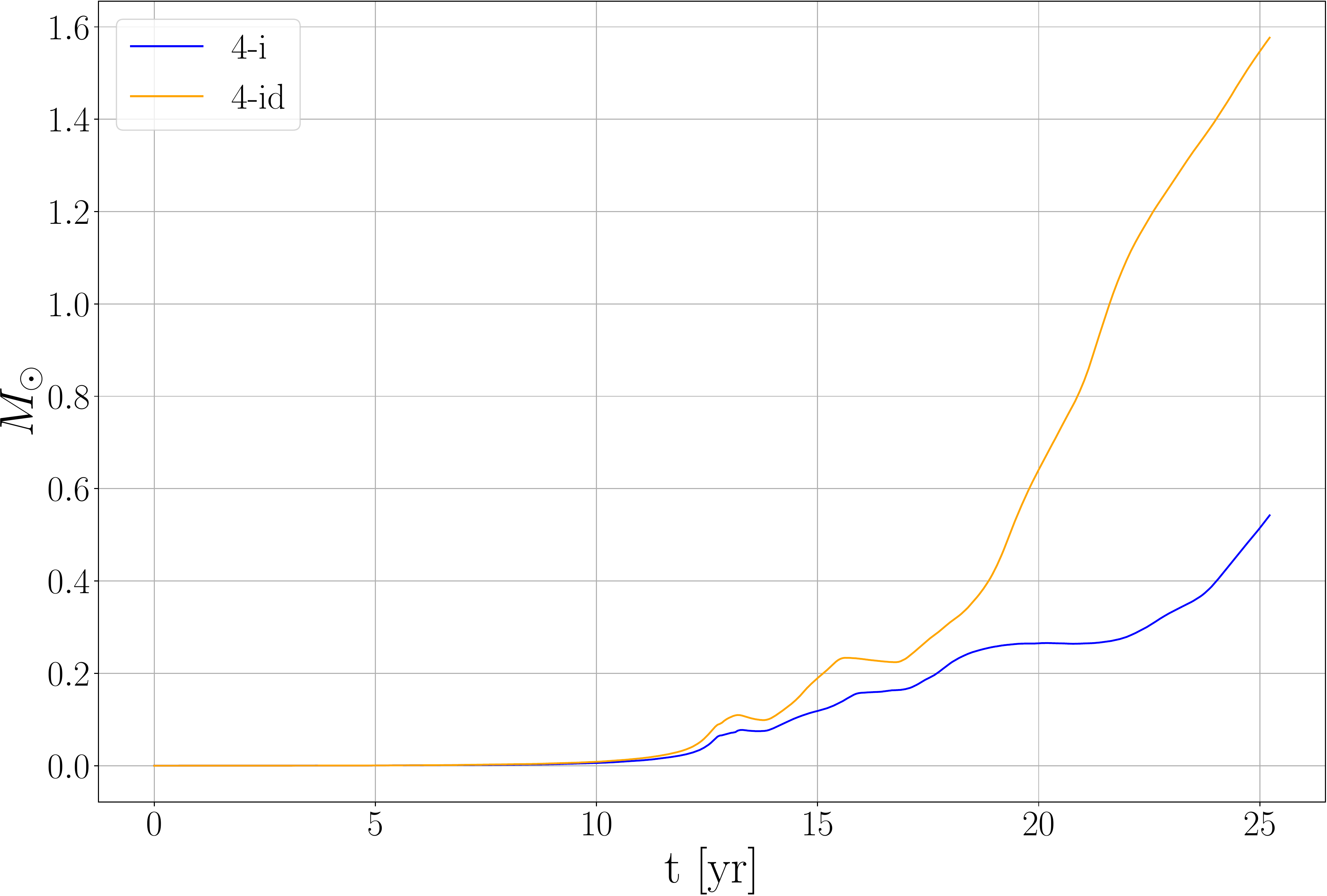}
         \end{subfigure}
     \hfill
     \begin{subfigure}[b]{0.48\textwidth}
         \centering
         \includegraphics[width=\textwidth]{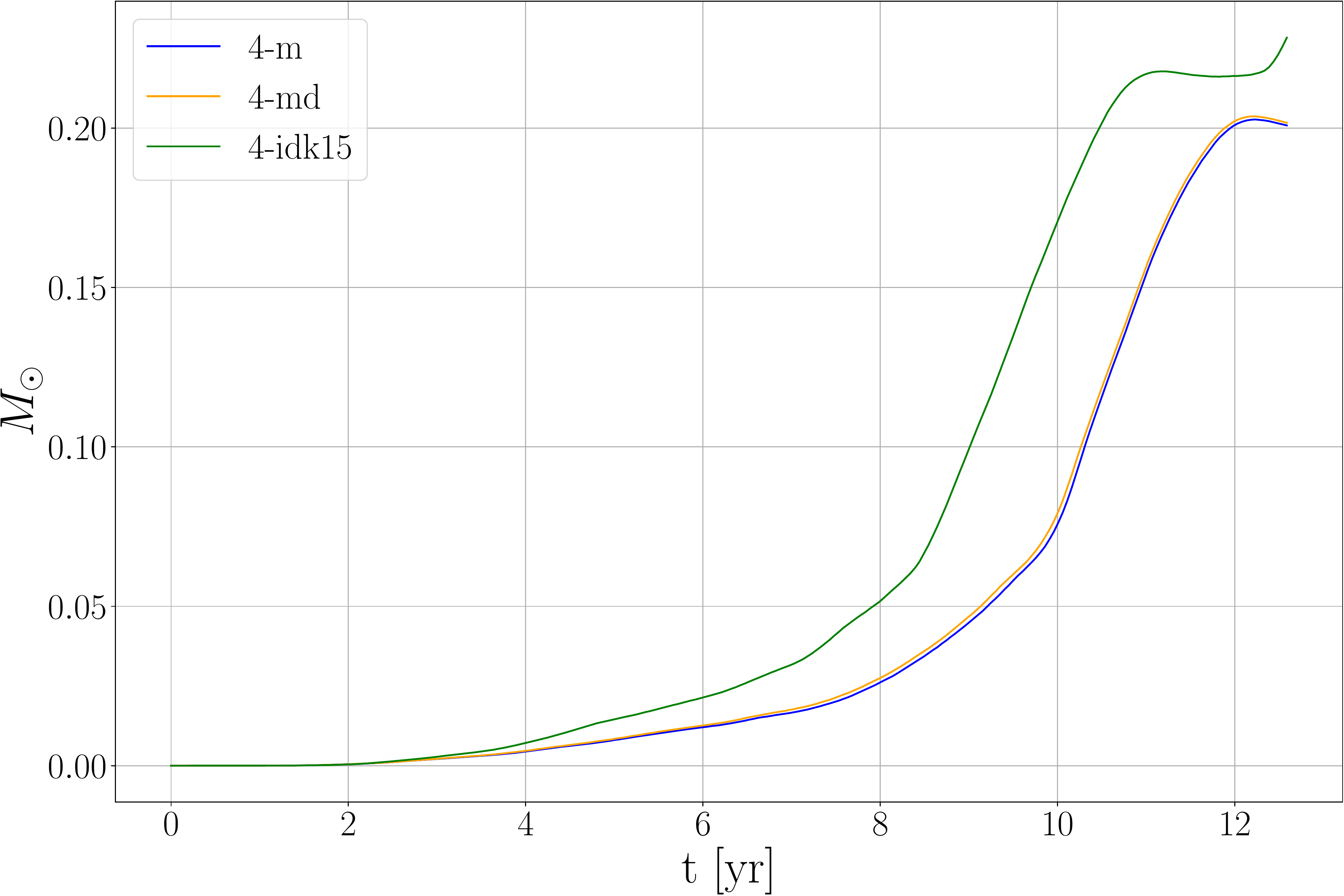}
         \end{subfigure}
     \hfill
     \centering
     \begin{subfigure}[b]{0.48\textwidth}
         \centering
         \includegraphics[width=\textwidth]{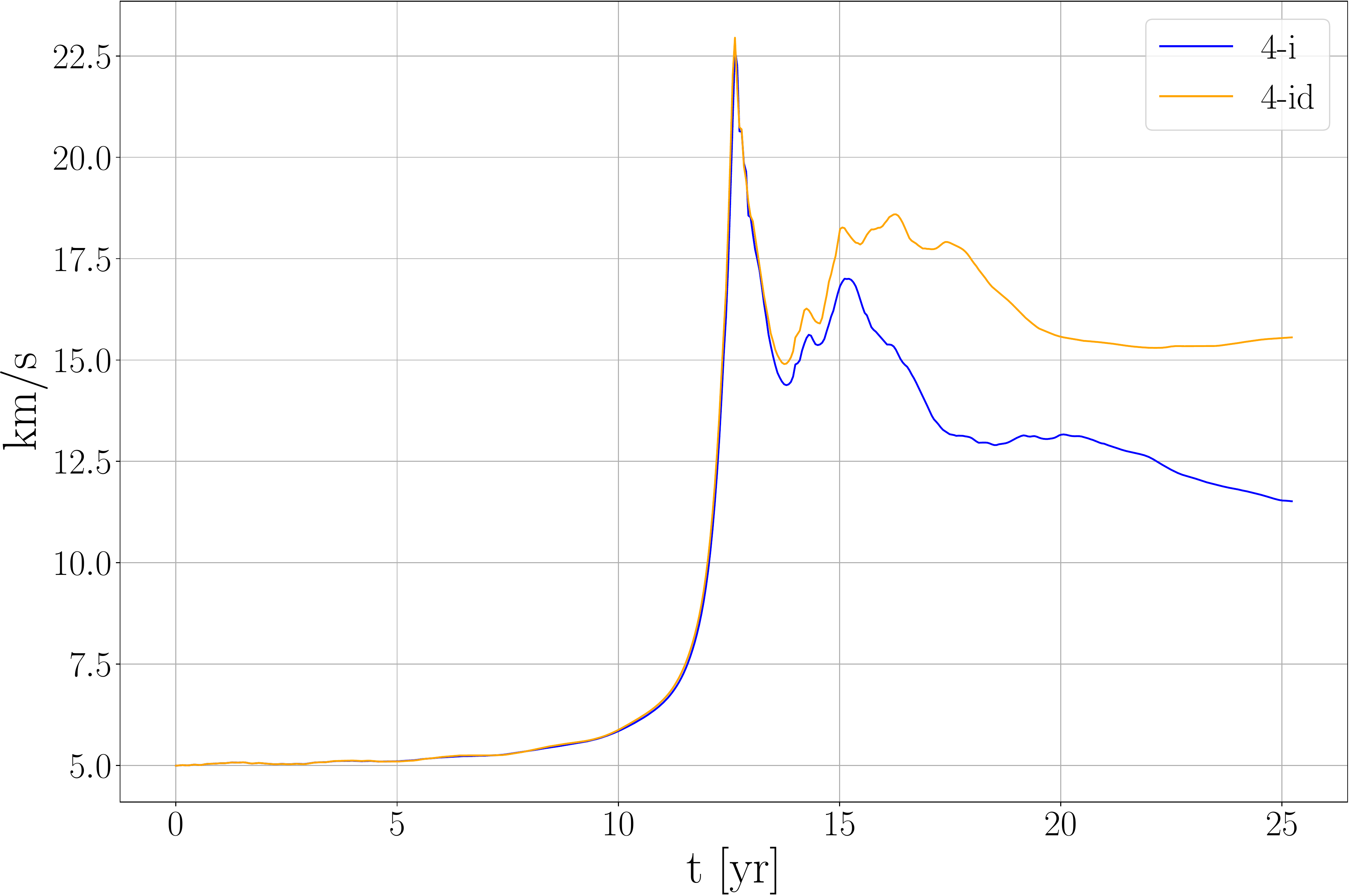}
         \end{subfigure}
     \hfill
     \begin{subfigure}[b]{0.48\textwidth}
         \centering
         \includegraphics[width=\textwidth]{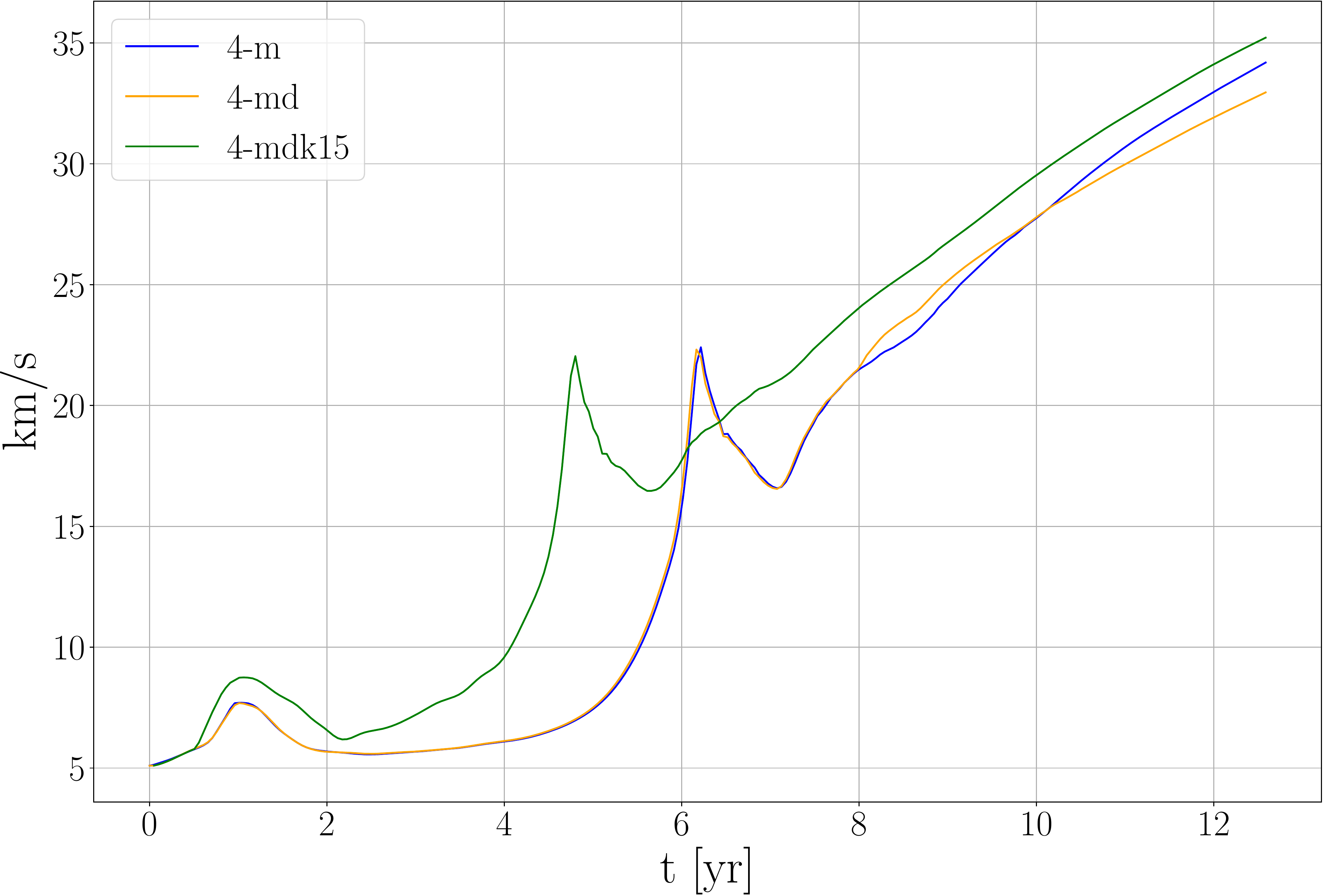}
         \end{subfigure}
     \hfill
     \caption{Top row: the total mass of SPH particles with $T \le 2000$~K for the 3.7~\Msun\ simulations. Bottom panels: the average speed of the gas with T$\le 2000$~K for the same simulations. \review{Top right panel, green line shows label 4-mdk15 now, instead of 4-idk15.}}
    \label{fig:mv4M}
\end{figure*}

In Figures~\ref{fig:mv2M} and \ref{fig:mv4M}  we show the effect of the Bowen implementation on the linear momentum of the 1.7 and 3.7~\Msun\ models by plotting the mass and the average speed of the gas with $T \le T_{\mathrm{cond}}$ for simulations with no dust acceleration and for two simulations with dust accelerations with $\kappa_{\rm max} = 5$ and 15. Left panels are for the ideal gas simulations and the right panels are for the tabulated EoS simulations. 

 In the 1.7~\Msun\ model (top-left panel, Figure~\ref{fig:mv2M}), dusty and non dusty curves separate at 6 yr, a consequence of the earlier onset of plunge-in in the dusty models (Fig.~\ref{fig:boundsep}). Even if we shifted the dusty model curves forward to synchronize the time of in-spiral, we would still see that dusty models produce more cool mass; we conclude that dusty models suffer more expansion and adiabatic cooling, even if as we saw in Section~\ref{ssec:orbital_evolution} and Figure~\ref{fig:boundsep} this does not lead to a great deal more unbound mass. As for the tabulated EoS models, the amount of cool mass is barely changed with the inclusion of dust. 


The bottom plot in Figure~\ref{fig:mv2M} is the average speed of the cool gas for the 1.7~\Msun\ models. While the 2-i and 2-id curves have similar behaviours, the vertical shift between them goes from $3$~km~s$^{-1}$ at $t~\approx~1.5$--$2$ yrs to $\sim10$ km~s$^{-1}$ at the time of steepest in-spiral. We included 2-idk15 but remind the reader that this model was started using an output of the 2-id simulation taken at $\sim$7.5~yrs. Even so, increasing  $\kappa_{\mathrm{max}}$ produces an immediate rise of the average velocity and incidentally of the cool mass. On the other hand, the inclusion of recombination energy (right panels, Figure~\ref{fig:mv2M}) does not change significantly the flow velocity: the higher internal energy in these tabulated EoS  simulations leads to a higher gas temperature. While this does not increase the surface temperature by a wide margin, it speeds up the CE interaction and reduces the time window for the dust opacity to accelerate the gas.     

In Figure~\ref{fig:mv4M} we show a similar analysis for the 3.7~\Msun\ simulations. For the ideal gas EoS simulations, it is clear from the left panels that substantially more mass reaches cool temperatures in the dusty simulations, but this mass is not accelerated a great deal compared to the non-dusty simulations. This explains the almost identical in-spiral shape for dusty and non-dusty simulations. However, for the 3.7~\Msun\ simulations with recombination energy, we observe that the RLOF timescale is affected by the dust-driven acceleration, but only when $\kappa_{\rm max}=15$\kapunits. Even then, the difference in the cool mass is small and does not provide a significant amount of momentum to the gas.  



\subsection{Opacity distribution}
\label{ssec:opacity_distribution}

\begin{figure*}
    \begin{minipage}{0.48\textwidth}
        \centering
        \includegraphics[width=\textwidth]{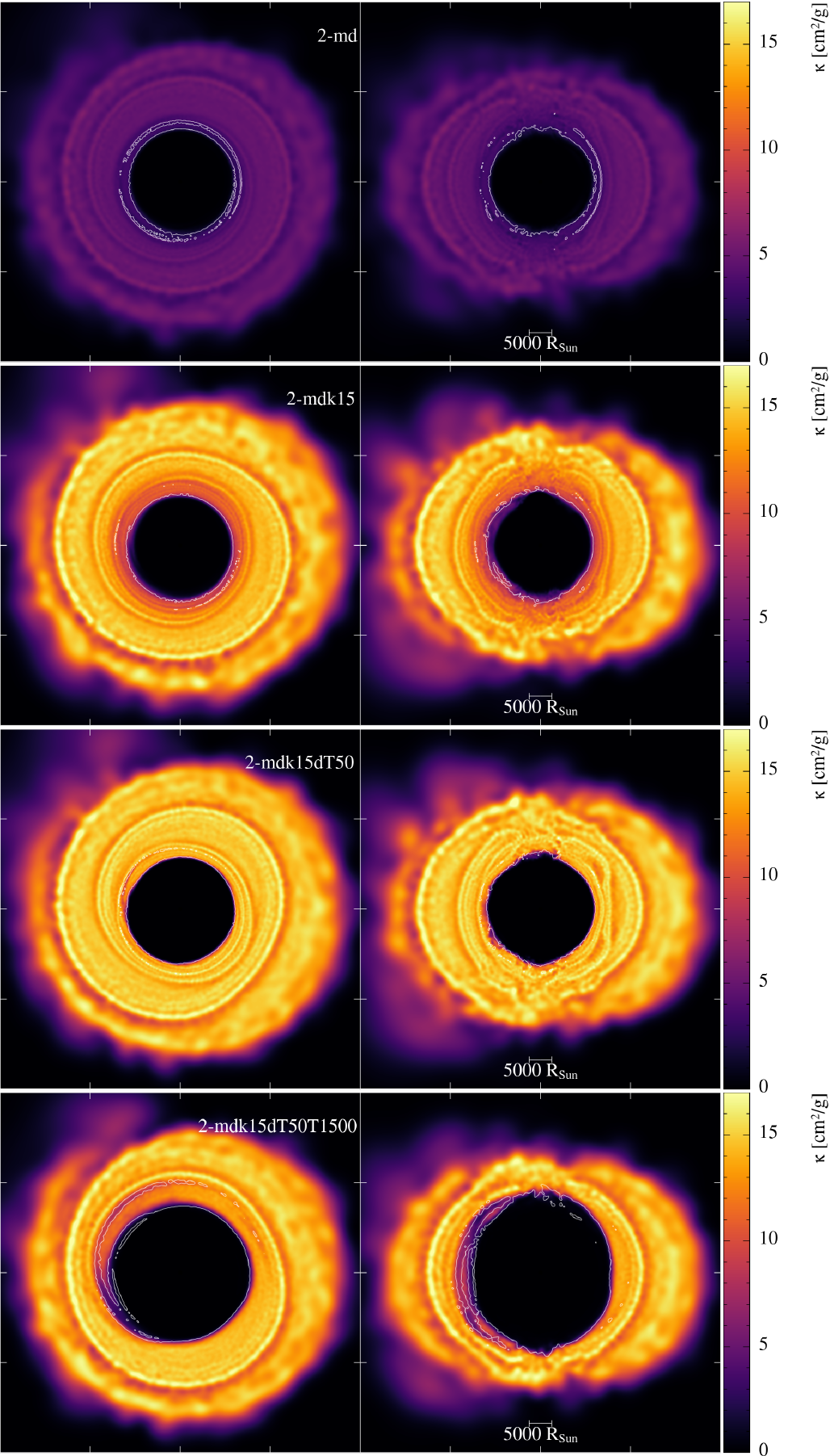}
        \subcaption{Simulations with the 1.7 \Msun donor star. First and second columns \\ show the orbital (x-y) and perpendicular (x-z) planes, respectively.}
        \label{fig:kappa_2M}
    \end{minipage}
    \hfill
    \begin{minipage}{0.48\textwidth}
        \includegraphics[width=\textwidth]{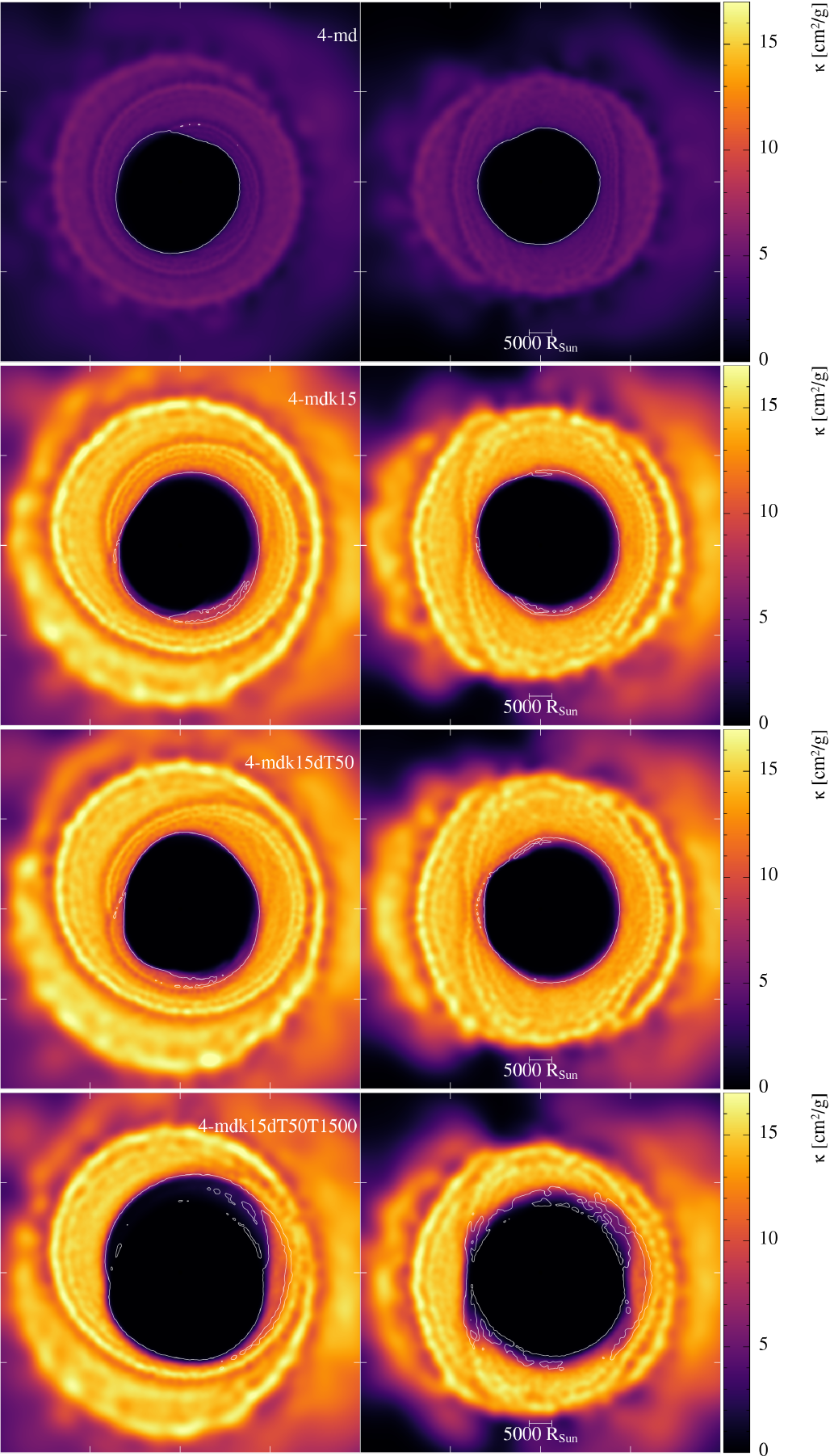}
        \subcaption{Simulations with the 3.7 \Msun donor star. First and second columns \\ show the orbital (x-y) and perpendicular (x-z) planes, respectively.}
        \label{fig:kappa_4M}
    \end{minipage}
    \hfill
    \caption{Opacity cross sections for the models with a tabulated equation of state and dust-driving. All panels were taken at the end of each simulation ($t=12.6$ years). Each rendered model is indicated in the top right panel of the first column in Figures~\ref{fig:kappa_2M} and \ref{fig:kappa_4M}. Opacity movies of each simulation are available at the following link: \url{https://miguelglezb.github.io/mgb/simulations_page/bowen-dusty-ce.html}.}
    \label{fig:kappa_render}
\end{figure*}

In Figure~\ref{fig:kappa_render}, we show slices of opacity for the eight models with a tabulated EoS and Bowen dust. The left set of four simulations is for the 1.7~\Msun\ models, while the right set is for the 3.7~\Msun\ models. The first row is for the `md' models with default values: $T_\mathrm{cond}=2000$~K, $\delta=200$~K and $\kappa_{\rm max}=5$\kapunits. The second row shows the `mdk15' models, which have the same values as the previous model, but with an increased $\kappa_{\rm max} = 15$\kapunits. The third row displays the `mdk15dT50' models, which have the same $\kappa_{\rm max}$ value as the previous models, but have a decreased $\delta$ = 50~K. Finally, the last row represents the `mdk15dT50T1500' models, which have the same values as the previous models, but with a decreased T$_{\rm cond}=1500$~K. 

For both 1.7 and 3.7~\Msun\ models, increasing $\kappa_{\rm max}$ increases the overall opacity of the model, but does not really impact the morphology of the flow. Decreasing $\delta$ has only a minor effect, slightly increasing the opacity, mainly in the central regions, but not enough to warrant consideration. Decreasing the T$_{\rm cond}$ value, on the other hand, increases the radius of the inner boundary for the high opacity region. This decrease in optical depth close to the star may decrease the amount of gas that is accelerated outwards by dust-driving. Yet this does not appear to have a large effect on the in-spiral and unbound mass \review{for models with recombination energy} (Fig.~\ref{fig:boundsep}).

%% file: discussion/discussion_conclusions.tex
\section{Summary and discussion}
\label{sec:conclusion}

\begin{enumerate}
    \item The inclusion of dust-driven accelerations tends to hasten the CE in-spiral by increasing the rate of expansion of the outer layers at early times, i.e., during the RLOF phase (the only model for which this is not so is the ideal gas EoS, 3.7~\Msun\ simulation).

    \item The final orbital separation is $\sim10-25$~per cent larger with dust-driven accelerations for the 1.7~\Msun\ models (for the 3.7~\Msun\ models we cannot draw any conclusion because the final separations are similar to the softening length of the core). 

    \item \review{Eccentricity in post-CE systems depends on the interaction between the envelope and the companion during the plunge-in. If the gas takes more angular momentum, relative to the final orbital energy of the point mass particles, the final eccentricity will increase. In general, we find larger eccentricities in simulations with dust-driven acceleration. This suggests that the (typically) higher gas velocities in dusty simulations transfer more angular momentum to the ejected gas, leading to more eccentric orbits of the point mass particles at later times.}
    
    \item The 1.7~\Msun\ dust-driven simulations unbind $\sim5$~per cent more gas in the ideal gas EoS scheme. However, they unbind $\sim10$~per cent {\it less} than the non-dusty model with recombination EoS. Some of these differences may be due to timing.  The 3.7~\Msun\ dusty simulations unbind $\sim5$~per cent more mass with some of that difference also due to the timing of the in-spiral at early times.
    
    \item In general it seems that simulations that include the deposition  of recombination energy are less affected by the inclusion of dust-driving. Recombination energy overwhelms the relatively small effect of the dust-driven wind acceleration, because the thermal energy deposited in the gas by recombination already produces relatively large accelerations. 
    
    \item The accelerations due to dust-driving are larger along the equator than along the poles, but only for the simulations with an ideal gas EoS. For simulations that include recombination energy the deposition of additional thermal energy when the gas recombines pushes the gas more isotropically, with only a small hint of equatorial enhancement, particularly for the more massive, 3.7~\Msun\ simulation. This would imply that the presence of dust does not, {\it per se}, lead to a strongly bipolar morphology.

    \item Compared to non dusty models the photosphere of the CE object is much larger. At the end of the both 1.7~\Msun\ and 3.7~\Msun\ simulations (12.5~yr) the dusty photosphere is $\sim$~10 times larger than the corresponding non-dusty model.


\end{enumerate}

In conclusion we simulated dust driving using opacities calculated with the Bowen approximation for  two  models of AGB stars with masses of 1.7 and 3.7~\Msun\ and companions of 0.6~\Msun. For these two models, the dust opacity is more important for the optical properties of the models with a clear effect on the size of the photosphere, rather than for driving the envelope. In Paper~II we will consider the same questions, but with a more realistic calculation of the dust opacities that uses a full nucleation network. We expect that as gas expands and cools adiabatically, dust will start to nucleate. However, dust nucleation will not be instantaneous, so that the dust opacity will not grow immediately as is instead the case with the Bowen approximation. This may lead to some differences both in the dynamical and optical properties of the common envelope.

%% file: appendices/stellar_relaxation.tex
\section{Stellar relaxation}
\label{app:stellar_relaxation}

In this Appendix we describe the stellar relaxation procedures for the 3.7~\Msun\ stellar profile using the technique developed by \cite{Lau2022a}. The relaxation procedure is the same as described by \cite{GonzalezBolivar2022}: after mapping the \mesa\ 1D profile into the 3D computational domain the star is relaxed. Following the relaxation procedure the star is tested in isolation by evolving it for a number of dynamical times, at the end of which we evaluate the degress of expansion and the velocities that may have developed.

For the 1.7~\Msun, \mesa\ EoS models, we have used the same stellar model as that of  \cite{GonzalezBolivar2022}. However, for the binary simulations carried out in this paper, we have used as input a  stellar model taken at a later time in the isolated star run, compared to their original binary simulation (see Appendix C of \cite{GonzalezBolivar2022}). In their case, the input stellar model was taken 0.6 yr after the end of relaxation. In this work, we take it at 2.4 yrs after the end of relaxation, when the radius of the star has stabilised considerably, albeit at a slightly larger dimension \citep[see figure 6 of][and relevant text]{GonzalezBolivar2022}. 

In Figure~\ref{fig:rho-R} we show a similar plot to that found in appendix A of \cite{GonzalezBolivar2022}. The first column is the 1.7~\Msun\ model, right after mapping, after relaxation and 3 years after relaxation ends (top, middle and bottom rows, respectively; the last is the model used as input for the CE simulations). The second and last columns are equivalent plots for the 3.7~\Msun\ models with an ideal gas and a tabulated EoS, respectively. The input models (bottom row) for the 3.7~\Msun\ star, are taken 0.6 years after the end of relaxation.

Isolated stellar models with a 3.7~\Msun\ star and an ideal gas or tabulated EoS (second and third columns in Figure~\ref{fig:rho-R}) have similar characteristics compared to their low mass counterparts. The ideal gas EoS with a 3.7~\Msun\ star is the most stable model of all, which means it barely expands when run in isolation. The 3.7~\Msun\ model with a tabulated EoS is, like its 1.7~\Msun\ counterpart, less stable. That is why there is a significant expansion of the star after relaxation (third row, Figure~\ref{fig:rho-R}). Both 3.7~\Msun\ models were run for 5 dynamical times ($\sim$200 days) after the relaxation was turned off and the difference in the expansion is visible in the bottom row panels (central and right columns) of Figure~\ref{fig:rho-R}.   

\begin{figure*}
     \centering
     \includegraphics[width=0.8\textwidth]{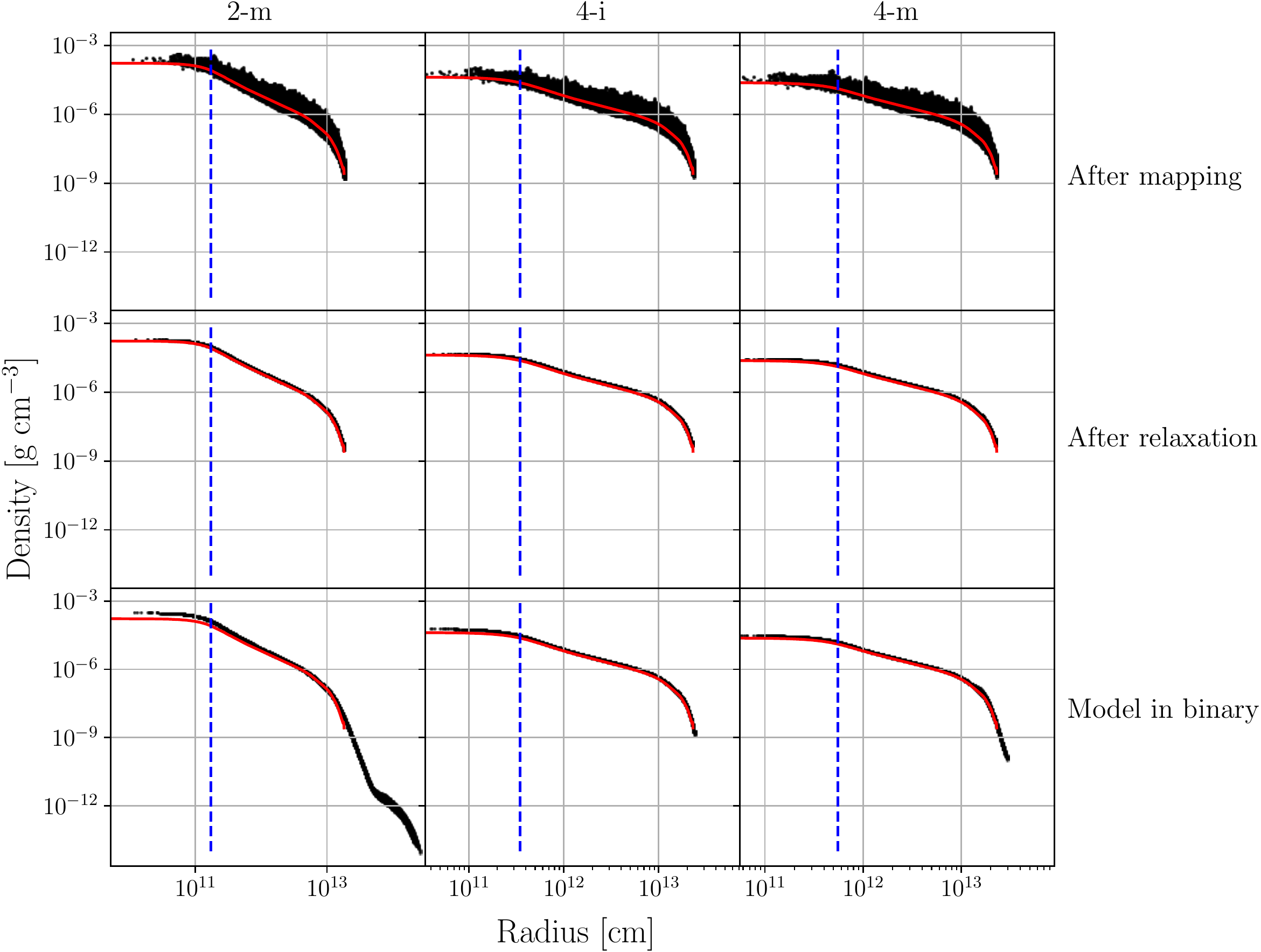}
     \caption{Density profile models new in this paper (model 2-i was already presented by \citet{GonzalezBolivar2022}, while model 2-m is similar to the one presented before, but the third panel is taken at a later time (see text)). The mapped profiles (top panels) display a thick distribution of SPH particles, because the initial distribution is affected by the discretization of the original stellar density profile (shown as red curves). Second row: profiles right after the relaxation procedure. Bottom row:  the model evolved in isolation for 15 and 5 dynamical times for the 1.7 and 3.7~\Msun\ cases, respectively. The dashed blue vertical lines indicate the softening length of the point mass particle core of each density stellar profile. }
     \label{fig:rho-R}
\end{figure*}

\begin{figure}
     \centering
     \includegraphics[width=0.4\textwidth]{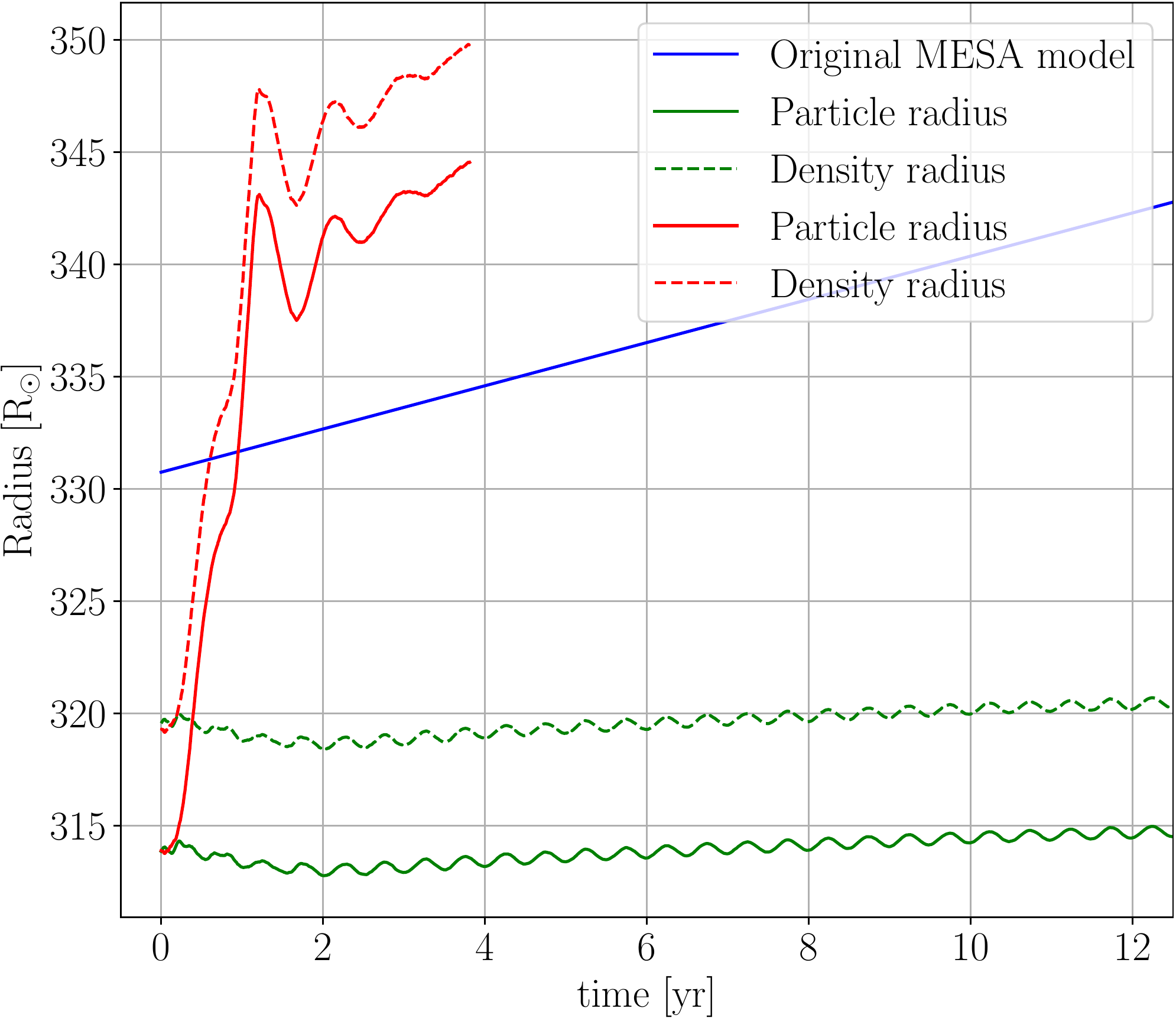}
     \caption{\review{Radial evolution of the isolated 4-i (green) and 4-m (red) stellar models.}}
     \label{fig:PD_rad}
\end{figure}

\review{We have compared the radial expansion of the 3.7~\Msun\ stellar model carried out by the \mesa\ code beyond the profile used in the common envelope simulations, as well as the \phant\ 3D models, but run in isolation (Figure~\ref{fig:PD_rad}). Given the nature of the SPH particles, measuring stellar radii is not a straightforward endeavour. Therefore, we adopted the ``particle" and ``density" radii criteria also used by \cite{GonzalezBolivar2022} for the 1.7~\Msun\ donor star. The ``particle" radius is the average of the radius of the farthest 0.5~per cent of all SPH particles of the model. The ``density" radius is defined as the volume-equivalent radius \citep{Nandez2014} of all the SPH particles whose density is higher than the least dense SPH particle at $t=0$. For the latter case, $t=0$ is the time of the models that are shown in the bottom (center and right) panels in Figure~\ref{fig:rho-R}. Three models were run in isolation with no relaxation procedure. At first, the radii of the 3D models are smaller due to the contraction of the star during the precedent relaxation. Model 4-i remains at a lower radius compared to the stellar evolution in \mesa. However, model 4-m shows a radial expansion of $\approx$30 \Rsun\ per year during the first year, reaching values of 347 and 343\Rsun\ for density and particle radii, respectively. Model 4-m was halted at $t\approx 4$ years due to a decrement of the Courant time-stepping of two orders of magnitude compared to the initial time, making the model too computational expensive to keep running.}

\begin{figure}
     \centering
     \includegraphics[width=0.42\textwidth]{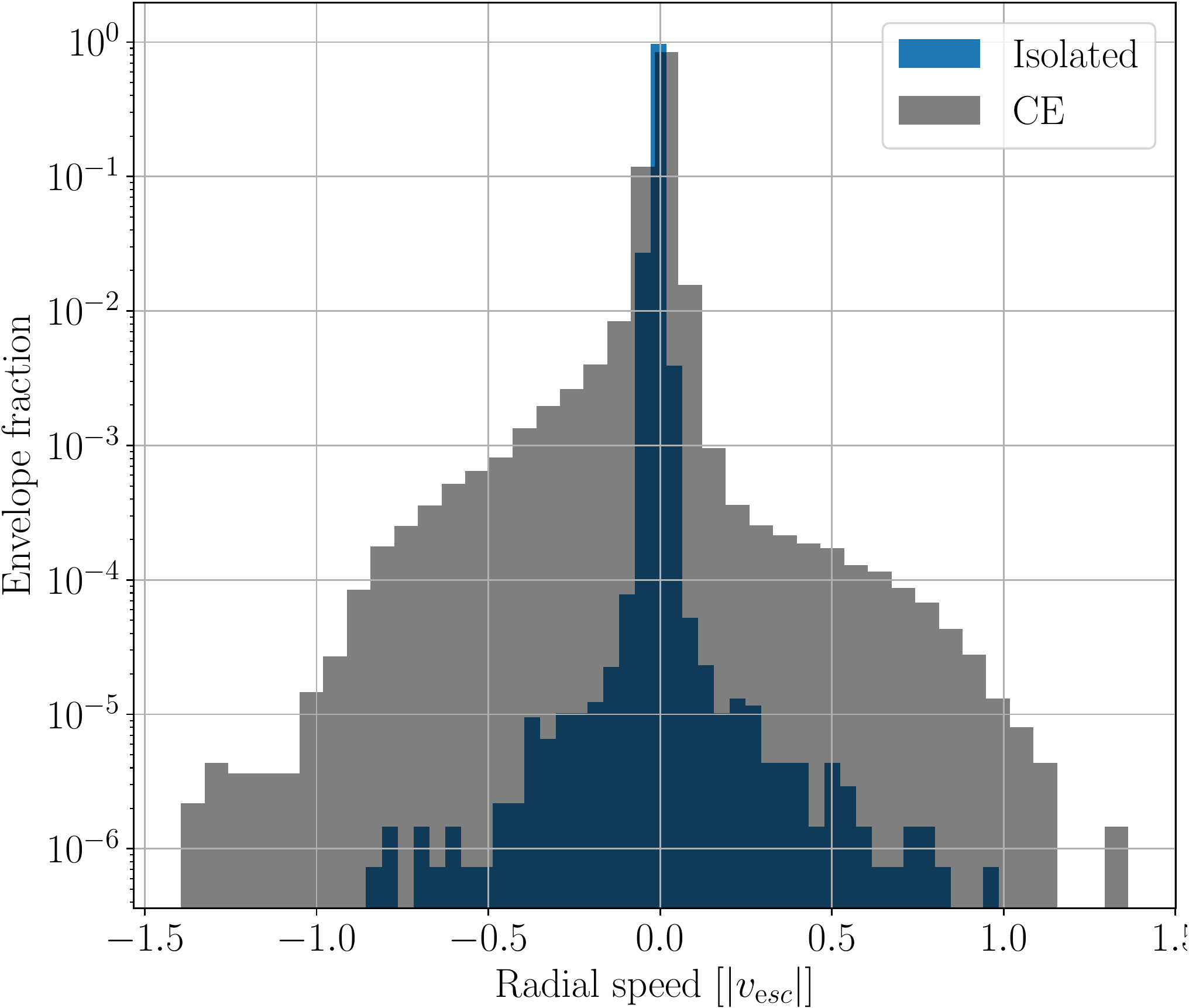}
     \caption{\review{Radial speed distribution with respect of the giant core's point mass particle for the 4-m model in the common envelope (gray) and isolated (blue) simulations at $\approx 4$ years. The radial speed is indicated in units of escape velocity ($\approx 64$km~s$^{-1}$) from the donor star.}}
     \label{fig:vel_hist}
\end{figure}

\secondreview{Given the expansion of the original \mesa\ profile, and as a sanity check of the stability of the model, we analysed the velocity distribution of the 3.7\Msun\ with and without the companion at $t=4$ years (Figure~\ref{fig:vel_hist}). The velocity distribution of the isolated model (blue) shows that the envelope is expanding, but it is far from reaching the escape velocity of the star. Conversely, a fraction of the envelope is being ejected when is set in orbit with the companion (gray) and evolved for the same amount of time.}

%% file: appendices/computational_resources.tex
\section{Computational resources and wall clock time}
\label{app:comp_resources}

The binary simulations were carried out on the Gadi supercomputer of the National Computational Infrastructure (NCI), using a single node of 24 processors (48 cores) Intel(R) Xeon(R) Platinum 8268 CPU 2.90GHz (models 2-i, 2-m and all the 3.7~\Msun\ models);
on a virtual machine (VM) granted by the ``Oracle for Research" programme, using a 80 Neoverse-N1 cores (models 2-idk15 and 2-mdk15); and on an in-house server with 48 Intel(R) Xeon(R) Platinum 8168 CPU 2.70GHz processors (model 2-md). Model 2-id was run from t=0 to t=12.51 yrs on the in-house server and the remaining simulation time was completed on the Oracle VM. 

Simulations run with \phant\ version 2022.0.1, particularly 4-m and 4-md, have wall-clock times of 10.2 and 10.8 days, respectively. Since the Bowen implementation just requires a slight modification in the momentum treatment of the gas (Equation~\ref{eq:dvdt}) via a simple calculation of opacity (Equation~\ref{eq:kappa(T)}), it is expected that models with dust will have virtually no increase in wall clock time compared to their non-dusty counterparts. Simulations 4-i and 4-id, on the other hand, have wall-clock times of 20 and 15 days, respectively. This decrease in wall clock time was caused by the depletion of gas near the point mass particles, which have typically the smallest Courant time step. Therefore, the time step increases because of the ejection of material from the central region. This effect is not seen in simulations with recombination energy, because the higher temperatures, compared to the models with an ideal gas EoS, near the point mass particles inhibits the dust acceleration of the gas. Hence, the Bowen implementation will have either virtually the same wall clock time or a decrement compared to models without dust.   

Given that the 1.7~\Msun\  simulations with dust were carried out with a more recent, better optimised, versions of \phant\ relative to their non-dusty counterpart models, we cannot fairly compare their wall clock times.

